\documentclass[reprint,aps,pra,twocolumn]{revtex4-1}

\usepackage{bm}
\usepackage{graphicx}
\usepackage{times}
\usepackage{amsmath}

\begin{document}
\title{Mechanisms of optical orientation of an individual Mn$^{2+}$ ion spin in a II-VI quantum dot}

\author{T. \surname{Smole\'nski}$^{1}$}\email{Tomasz.Smolenski@fuw.edu.pl}
\author{\L. \surname{Cywi\'nski}$^{2}$}
\author{P. \surname{Kossacki}$^{1}$}

\affiliation{
$^{1}$  Institute of Experimental Physics, Faculty of Physics, University of Warsaw, ul. Pasteura 5, 02-093 Warsaw, Poland \\
$^{2}$  Institute of Physics, Polish Academy of Sciences, Al. Lotnik\'ow 32/64, 02-688 Warsaw, Poland}

\date{\today}

\begin{abstract}
We provide a theoretical description of the optical orientation of a single Mn$^{2+}$ ion spin under quasi-resonant excitation demonstrated experimentally by Goryca \textit{et al.} [Phys. Rev. Lett. \textbf{103}, 087401 (2009)]. We build and analyze a hierarchy of models by starting with the simplest assumptions (transfer of perfectly spin-polarized excitons from Mn-free dot to the other dot containing a single Mn$^{2+}$ spin, followed by radiative recombination) and subsequently adding more features, such as spin relaxation of electrons and holes. Particular attention is paid to the role of the influx of the dark excitons and the process of biexciton formation, which are shown to contribute significantly to the orientation process in the quasi-resonant excitation case. Analyzed scenarios show how multiple features of the excitonic complexes in magnetically-doped quantum dots, such as the values of exchange integrals, spin relaxation times, etc., lead to a plethora of optical orientation processes, characterized by distinct dependencies on light polarization and laser intensity, and occurring on distinct timescales. Comparison with experimental data shows that the correct description of the optical orientation mechanism requires taking into account Mn$^{2+}$ spin-flip processes occurring not only when the exciton is already in the orbital ground state of the light-emitting dot, but also those that happen during the exciton transfer from high-energy states to the ground state. Inspired by the experimental results on energy relaxation of electrons and holes in nonmagnetic dots, we focus on the process of biexciton creation allowed by mutual spin-flip of an electron and the Mn$^{2+}$ spin, and we show that by including it in the model, we obtain good qualitative and quantitative agreement with the experimental data on quasi-resonantly driven Mn$^{2+}$ spin orientation.
\end{abstract}

\maketitle

\section{Introduction}
Manipulation of single spins localized in semiconductor materials has been a focus of great attention both due to prospects of their use for quantum computation purposes \cite{Hanson_RMP07,Liu_AP10}, and also for other nano-electronic applications \cite{Koenraad_NM11}. Carrier (electron or hole) spins localized in quantum dots (QDs) can be coherently controlled electrically \cite{Hanson_RMP07} and, more importantly in the context of this paper, optically \cite{Liu_AP10,DeGreve_RPP13}. In fact, a single quantum dot can be selectively excited by laser light, and its photoluminescence can be detected \cite{Gammon_PRL96}, allowing for a direct measurement of a single nanoscale system.  

Spins of transition metal impurities (e.g.,~Mn in II-VI and III-VI materials) in diluted magnetic semiconductors (DMS) are strongly coupled to carriers by the $s,p$-$d$ exchange interaction, leading to well-known pronounced magneto-optical effects in II-VI materials \cite{Gaj_book_2010}, and to carrier-mediated ferromagnetism in II-VI~\cite{Boukari_PRL02} and III-V materials~\cite{Dietl_Science00,Dietl_NM10}. The optical orientation of ensembles of these spins (i.e.,~creation of their magnetization due to absorption of circularly polarized light) was observed almost thirty years ago in (Hg,Mn)Te \cite{Krenn_PRL85,Krenn_PRB89} and (Cd,Mn)Te \cite{Awschalom_PRL85}. Later, precession of photo-oriented Mn$^{2+}$ spins in transverse magnetic field was observed in (Cd,Mn)Te quantum wells \cite{Crooker_PRB97}. More recently, this phenomenon was also optically detected for a relatively small number (a few hundred) of Mn spins in (Ga,Mn)As quantum well \cite{Myers_NM08}. It should be stressed that for the main Mn spin relaxation mechanisms, both transverse and longitudinal, are made possible by short-range Mn-Mn exchange interactions, the importance of which grows with increasing Mn concentration. Since the orbital moment of Mn$^{2+}$ ion (which has a half-filled shell with spin $S=5/2$) is zero, the spin-orbit interaction effects are strongly suppressed, so that the direct process of energy relaxation via phonon emission is expected to be weaker than relaxation processes made possible by Mn-Mn interactions. The longitudinal relaxation of Mn spins in (II,Mn)VI materials is usually explained by phonon-induced modulation of Mn-Mn interactions, \cite{Scalbert_pssb96} with the hyperfine interaction between the Mn electronic spin and its nuclear spin starting to play a role at very small Mn concentrations and at magnetic fields close to zero \cite{Goryca_relaxation_PRL09}. Transverse Mn spin relaxation is also caused by anisotropic part of Mn-Mn superexchange \cite{Samarth_PRB88,Larson_PRB89}, which also explains the dynamics of  magnetization creation in a magnetic polaron \cite{Dietl_PRL95,Klopotowski_PRB11}.

It could be thus expected that an isolated Mn$^{2+}$ spin should have long relaxation and coherence times, while the $s,p$-$d$ exchange interaction should allow for its manipulation with the help of electrons and holes, controlled either optically or electrically. These expectations spurred intense research on QDs containing single transition metal ions. After the first optical measurements identifying CdTe/ZnTe dots containing single Mn$^{2+}$ spins \cite{Besombes_PRL04}, many features of the energy spectrum and the dynamics of single Mn spins in CdTe/ZnTe QDs \cite{Besombes_PRB05,Leger_PRL06,Leger_PRB07,Besombes_PRB08,LeGall_PRL09,Goryca_PRL09,LeGall_PRB10,Goryca_PRB10,LeGall_PRL11,Trojnar_PRL_2011,Trojnar_PRB_2013,Varghese_PRB_2014,Goryca_PRL_2014,Lafuente_PRB_2015,Smolenski_PRB_2015_x2m}, CdSe/ZnSe QDs \cite{kobak_nature_2014,Smolenski_PRB_2015,Smolenski_JAP_2015} and InAs/GaAs QDs \cite{Kudelski_PRL07,Krebs_PRB09,Baudin_PRL11} were experimentally investigated. Among others, this includes demonstration of the optical orientation \cite{LeGall_PRL09,Goryca_PRL09,LeGall_PRB10,Baudin_PRL11,Smolenski_PRB_2015}, optical Stark effect \cite{LeGall_PRL11}, and coherent precession of a single Mn$^{2+}$ spin \cite{Goryca_PRL_2014,Lafuente_PRB_2015}, as well as determination of its relaxation time in different QD systems \cite{Goryca_PRL09,kobak_nature_2014,Goryca_PRB_2015}. More recently, the research on magnetic-ion-doped QDs was also extended to the dots containing other transition metal ions, such as Co$^{2+}$ \cite{kobak_nature_2014}, Fe$^{2+}$ \cite{Smolenski_Nature_2016} and Cr$^{2+}$ \cite{Lafuente_PRB_2016,Lafuente_APL_2016} exhibiting non-zero angular momentum and thus being very sensitive to the local strain.

The key optical feature of a QD containing a single magnetic ion is a multifold splitting of the neutral exciton (X) emission line. In the particular case of a CdTe/ZnTe QD with a single Mn$^{2+}$, the photoluminescence (PL) spectrum consists of six strong lines corresponding to bright excitons shifted in energy by the exchange interaction (mostly $p$-$d$ hole-Mn$^{2+}$ interaction) with the six possible states of the Mn$^{2+}$ spin. For our considerations here it is crucial to note the presence of other, weaker lines, corresponding to mostly dark exciton states \cite{Leger_PRB07,Goryca_PRB10}, which are ``brightened'' by flip-flop parts of the $s$-$d$ and $p$-$d$ exchange interactions, the latter being allowed by the presence of significant heavy-hole (hh) - light-hole (lh) mixing in II-VI QDs. The magnitude of this hh-lh mixing can be estimated from the measurement of linear polarization of charged trion emission \cite{Koudinov_PRB04,Leger_PRB07}.

In this paper we focus on the optical orientation of a single Mn$^{2+}$ ion spin in a CdTe/ZnTe QD. In general, such an orientation can be induced by spin-polarized excitons that might be introduced to the QD under circularly-polarized optical excitation. We would like to stress here a somewhat underappreciated fact, that the exact microscopic mechanism of such a process for a single Mn$^{2+}$ spin, in the case of three-dimensional confinement of carriers that are coupled to it, is by no means obvious. In the case of bulk materials and quantum wells, continuous spectrum of carrier states of two spin polarizations was leading to existence of typically sizable phase space for spin-flip scattering between photo-excited and optically polarized carriers and the localized Mn$^{2+}$ spins \cite{Konig_PRB00}. Furthermore, as explained previously, the intrinsic (i.e.~not related to presence of photo-excited carriers) processes of spin relaxation of  Mn$^{2+}$ spins are becoming more efficient as the  Mn$^{2+}$ concentration increases, and in most of samples of dilute magnetic semiconductors the possibility of localized spins simply relaxing in the molecular field generated by spin-polarized photo-carriers is {\it a priori} possible (in reality one has to take into account competition between this process and the process of fast spin relaxation of carriers in presence of their large spin splitting generated by large number of  Mn$^{2+}$ ions interacting with each electron and hole \cite{Klopotowski_PRB11,Klopotowski_PRB13}). For a single Mn$^{2+}$ spin interacting with excitons confined in three dimensions, i.e.~the case of interest here, these simple mechanisms obviously do not work.

Sizable optical orientation of a single Mn$^{2+}$ spin was, however, demonstrated in two kinds of experiments utilizing different excitation techniques. The first one relies on resonant pumping of certain excitonic state in the Mn-doped QD, either an excited higher-energy orbital state \cite{LeGall_PRL09}, or one of six exchange-split ground states of X-Mn$^{2+}$ complex \cite{LeGall_PRB10}. The second technique involves quasi-resonant excitation of Mn-doped QD through a spontaneously created, adjacent QD \cite{Goryca_PRL09}, and the mechanism of optical orientation of the Mn$^{2+}$ spin in this case is the topic of the investigation presented in this paper. The quasi-resonant excitation relies on the fact that the CdTe QDs show a tendency to form chain-like patterns of closely spaced dots, some of which turn out to be strongly coupled by an efficient exciton tunneling process \cite{Kazimierczuk_PRB09}. Experimentally, such a coupled pair of QDs is identified as a sharp resonance in the photoluminescence excitation (PLE) spectrum \cite{Kazimierczuk_PRB09}. It occurs when the excitation laser energy is tuned to the ground state of the neutral exciton in a rather small QD, from which the exciton tunnels out in a few ps, and subsequently recombines from the orbital ground states of an adjacent larger QD, emitting light of energy about $0.2$~eV lower than the absorbed one \cite{Kazimierczuk_PRB09,Koperski_PRB_2014,Smolenski_PRB_2015_2X}. Importantly, the energy of resonant absorption of the smaller dot is independent of actual occupation a charge state of the emitting QD \cite{Kazimierczuk_PRB09,Smolenski_PRB_2016}, allowing for the latter dot to be occupied by multi-excitonic complexes similarly to the case of non-resonant above-the-barrier excitation. Nevertheless, the microscopic mechanisms responsible for creation of excitons in the QD in the two excitation regimes are qualitatively different. In fact, under non-resonant excitation the QD is occupied in a process of single-carrier trapping \cite{Suffczynski_PRB_2006, Kazimierczuk_PRB_2010}, while in the quasi-resonant regime the whole excitons are injected to the QD \cite{Kazimierczuk_PRB09}. Moreover, the quasi-resonant excitation is also shown to conserve the exciton spin orientation to a large degree of up to 70\% \cite{Kazimierczuk_PRB09}, allowing for efficient orientation of the Mn$^{2+}$ spin. In such an experiment, the absorbing dot is Mn-free, while the emitting dot contains a single Mn$^{2+}$ spin. The polarization-resolved PL measurements of this system revealed the occupation of all the six Mn$^{2+}$ spin levels to change upon excitation with circularly-polarized light, reaching the steady state characterized by a finite Mn$^{2+}$ spin polarization with $\langle S_z\rangle>0$ ($\langle S_z\rangle<0$) for $\sigma^-$ ($\sigma^+$) polarized excitation~\cite{Goryca_PRL09}.

From the theoretical side, only the process of the Mn$^{2+}$ spin orientation in the strictly resonant regime was quantitatively understood. In particular, it was shown that this orientation is due to the relaxation of the hole spin, which leads to the formation of a dark exciton in Mn-doped QD~\cite{Cywinski_PRB10,Cao_PRB11}. The key assumption making this orientation mechanism feasible is the possibility of undisturbed radiative recombination of the mostly dark state occurring with the change of the ion spin. However, this assumption is not valid under quasi-resonant excitation. More specifically, here the rate with which the excitons are injected to the emitting Mn-doped QD is independent of the actual occupancy of this dot, thus being governed solely by the excitation power. As a result, if the dark exciton resides in Mn-doped QD, it will not recombine radiatively even for relatively low excitation powers. Instead, the second exciton will be injected leading to the formation of a biexciton~\cite{Smolenski_PRB_2015_2X}. Thus, the observed efficient optical orientation of single Mn$^{2+}$ spins in quasi-resonant experiments with coupled dots needs to be explained.

Here we provide a detailed theoretical treatment of the possible processes leading to optical orientation of the Mn$^{2+}$ spin driven by the quasi-resonant excitation process described above. We start from a minimal set of assumptions about exciton transfer from the light-absorbing dot into the orbital ground state of the light-emitting one, and consider various theoretically possible mechanisms leading to optical orientation of the Mn$^{2+}$ spin. We discuss the character of the orientation that occurs both in the absence, and in the presence of carrier spin relaxation of the excitons. Crucially, we take into account the creation of biexcitons in the Mn-containing dot. The analysis of many increasingly complex models leads us to the main conclusion, that in order to describe the experimental results from \cite{Goryca_PRL09}, one has to take into account processes of carrier-Mn$^{2+}$ spin-flip that occur \emph{during} the relaxation of carriers towards the orbital ground state of the light-emitting QD. The mechanism that we propose leads to good agreement with observations, and it is analogous to the well-studied process of negative optical orientation of a negatively charged trion in a nonmagnetic QD~\cite{Cortez_PRL02,Kazimierczuk_PRB09,Benny_PRB_2014}.

The rest of the paper is organized as follows. The Hamiltonian of the system as well as the general assumptions behind the employed rate-equation model are described in Sec.~\ref{seq:model}. In Sec.~\ref{sec:minimal_model} we discuss possible mechanisms leading to the Mn$^{2+}$ spin orientation without any spin relaxation in the system. In particular, we demonstrate both the dynamics and direction of this orientation to be sensitively dependent on the QD parameters, e.g., the values of the $s,p$-$d$ exchange integrals. In Sec.~\ref{sec:dark_and_bright} we include in the model the possibility of the hole spin relaxation taking place during the exciton transfer between the coupled QDs, showing that in this case the spin orientation is faster, and its direction becomes independent of the model parameters. The comparison of the characteristics of the predicted orientation process with the ones known from the experiment~\cite{Goryca_PRL09} is provided in Sec.~\ref{seq:fullModel}. Based on this comparison we come to the conclusion that a correct explanation of the experiment requires including additional orientation mechanisms in the model, which would allow the Mn$^{2+}$ spin to be flipped during the exciton transfer between the dots. We show that a possible candidate for such a mechanism is the flip-flop of the electron and Mn$^{2+}$ spins occurring during the formation of a biexciton out of two excitons having parallel electron spins. We demonstrate that incorporation of such a mechanism in the model allows to quantitatively reproduce the experimental results from Ref.~\onlinecite{Goryca_PRL09}.

\section{The model}
\label{seq:model}
We consider the system of two laterally coupled CdTe QDs, one of them containing a single Mn$^{2+}$ ion. Laser light resonantly creates the excitons in the orbital ground state of the higher-energy Mn-free QD. The photo-created excitons tunnel out from this dot, and then undergo relaxation processes (energy relaxation and possibly spin relaxation) towards the orbital ground state in the lower-energy Mn-doped QD, from which the excitons recombine radiatively. Previous time- and polarization-resolved experimental studies of such QD pairs revealed that the tunneling time from the absorbing dot is of the order of a few ps \cite{Kazimierczuk_PRB09,Goryca_PRL_2014,Smolenski_PRB_2015_2X}, thus being much shorter than the radiative lifetimes of different bright excitonic complexes confined in the emitting Mn-doped QD (which are of the order of a few hundreds of ps \cite{Smolenski_PRB_2015_2X}). This allows us to assume that the rate of exciton injection to the emitting QD is independent of its actual occupancy and governed only by the excitation power. The second important experimental observation regarding the above-described quasi-resonant excitation is that the charge-state fluctuations of the emitting QD are much rarer than the events of an exciton injection \cite{Kazimierczuk_PRB09}. Thus, we will neglect these fluctuations in further analysis and focus solely on the neutral charge state of the Mn-doped QD. More specifically, we will consider both the states of the Mn$^{2+}$ ion in the empty QD and the states of X-Mn$^{2+}$ system. Importantly, for the complete description of the system it is also necessary to take into account the biexciton (2X) states. This is due to efficient formation mechanism of this excitonic complex from a long-lived dark exciton residing in the emitting QD, which was experimentally demonstrated to be responsible for significant 2X PL intensity of the emitting QD in the low-excitation-power regime \cite{Smolenski_PRB_2015_2X}. Naturally, for larger excitation intensities one should also consider higher excitonic states (e.g., the triexciton), however, here we neglect them and focus on the excitation powers remaining below the saturation of the X and 2X transitions.

\subsection{Hamiltonian of the QD with a single Mn$^{2+}$ ion}
\label{subseq:Model_Hamiltonian}

Let us first consider the Hamiltonian $\mathcal{H}_X$ of the exciton coupled to a single Mn$^{2+}$ ion in a QD. It is a sum of two terms $\mathcal{H}_{sp-d}$ and $\mathcal{H}_{e-h}$ corresponding to $s,p$-$d$ and electron-hole exchange interactions, respectively. In the description of both terms we use an effective model of an exciton living in a four-dimensional space spanned by spinful lowest-energy orbitals for an electron and a hole confined in a QD. The most important correction due to the presence of other orbitals is the existence of heavy-light hole mixing \cite{Koudinov_PRB04,Leger_PRB07}, and we take it into account by working with effective ``heavy hole'' states $|\Uparrow\rangle=|3/2\rangle+\sqrt{3}\epsilon|-1/2\rangle$ and $|\Downarrow\rangle=|-3/2\rangle+\sqrt{3}\epsilon^\dagger|1/2\rangle$, in which the complex parameter $\sqrt{3}\epsilon$ characterizes both strength and direction of the hh-lh mixing. Under this assumption the Hamiltonian $\mathcal{H}_{sp-d}$ can be written as \cite{Besombes_PRL04,Cywinski_PRB10,kobak_nature_2014,Smolenski_PRB_2015_x2m}
\begin{eqnarray}
\mathcal{H}_{sp-d} = &-&A_e\left[S_z s_z+\frac{1}{2}\left(S_+s_-+S_-s_+\right)\right]\\\nonumber
&+&A_h\left(S_zj_z+\epsilon S_+j_-+\epsilon^\dagger S_-j_+\right),  \label{eq:Hspd}
\end{eqnarray}
where $\vec{S}$ is the Mn$^{2+}$ spin operator ($S=5/2$), $\vec{s}$ is the electron spin operator ($s=1/2$), and $\vec{j}$ is the operator of an effective spin-$1/2$ acting in a two-dimensional space of effective hh states (with $\langle\Uparrow|j_z|\Uparrow\rangle=1/2$ and $\langle\Downarrow|j_z|\Downarrow\rangle=-1/2$). Parameters $A_e$ and $A_h$ are the effective exchange integrals of the Mn$^{2+}$ spin with the electron and the hole spin, respectively (note that the sign convention above is such that both of them are positive). In the simplest case of equal densities of the electron and hole wave functions at the Mn$^{2+}$ site in the QD, the ratio of these integrals $A_h/A_e$ should be equal to $|\beta/\alpha|\approx4$, where $N_0\alpha$ and $N_0\beta$ are the bulk (Cd,Mn)Te $s,p$-$d$ exchange constants \cite{Gaj_book_2010}. However, in the experiments on Mn-doped CdTe QDs the invoked ratio is typically slightly larger \cite{Besombes_PRL04,Smolenski_PRB_2015_x2m}, thus in the following we will take $A_h/A_e\approx5$.

\begin{figure}
\centering
\includegraphics{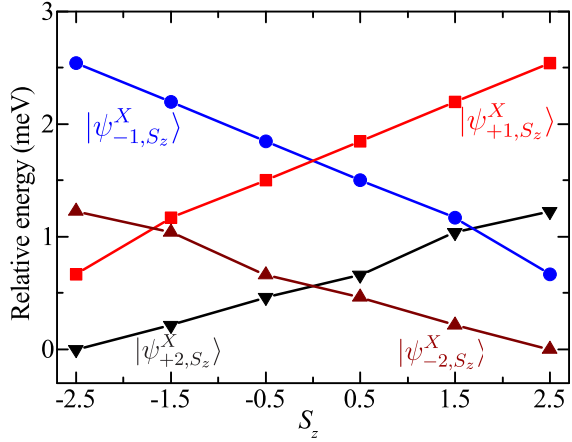}
\caption{Energy levels of the X-Mn$^{2+}$ system calculated based on numerical diagonalization of $\mathcal{H}_X$ Hamiltonian with an example set of parameters: $A_h=0.6$~meV, $A_e=0.12$~meV, $\epsilon=0.1$, $\delta_0=1.0$~meV, $\delta_1=0.1$~meV. The states $|\psi^X_{J_z,S_z}\rangle$ are displayed as function of their (relative) energy and dominant ion spin projection $S_z$. The color denotes the dominant exciton spin projection $J_z$ (as indicated).
 \label{fig0}}
\end{figure}

The second interaction defining the energy structure of the excitonic states is the electron-hole exchange, which can be described by the following Hamiltonian \cite{Bayer_PRB02}
\begin{equation}
\mathcal{H}_{e-h}=-2\delta_0s_zj_z+\frac{\delta_1}{2}\left(s_+j_-+s_-j_+\right),
\end{equation}
where $\delta_0$ is the isotropic exchange splitting between two bright states $\left|\Uparrow\downarrow\right\rangle$, $\left|\Downarrow\uparrow\right\rangle$ and two dark states $\left|\Uparrow\uparrow\right\rangle$, $\left|\Downarrow\downarrow\right\rangle$ (with $\uparrow$, $\downarrow$ representing the $+1/2$, $-1/2$ electron spin projections). In the following, we will label these states according to the projection of their total angular momentum as $|J_z=\pm1\rangle$ for the bright states and $|J_z=\pm2\rangle$ for the dark states (thus neglecting a small corrections due to the hh-lh mixing). The second term in the Hamiltonian $\mathcal{H}_{e-h}$ corresponds to the anisotropic part $\delta_1$ of the electron-hole exchange, which couples the two spin states of the bright excitons. In general, similar coupling arises also between the two dark X states \cite{Poem_NatPhys_2010,Gershoni_PRX_2015}, however we will neglect it in further considerations owing to its negligible strength $\delta_2\sim\mu$eV.

An example energy spectrum of the $\mathcal{H}_X=\mathcal{H}_{sp-d}+\mathcal{H}_{e-h}$ Hamiltonian is displayed in Fig.~\ref{fig0}. Its structure is mainly determined by three spin-conserving (Ising) terms $A_hS_zj_z$, $-A_eS_z s_z$ and $-2\delta_0s_zj_z$. They shift the energies of $|J_z,S_z\rangle$ excitonic states depending both on $J_z$ and $S_z$ leading to 12 doubly degenerate energy levels, half of which are bright, while the other half are dark. Within each of these subspaces, the energy levels are equally spaced with the splitting between the neighbouring states $|J_z,S_z\rangle$ and $|J_z,S_z+1\rangle$ being equal to $(A_h+A_e)/2$ or $(A_h-A_e)/2$ for the bright or dark states, respectively. Moreover, the dark states are shifted towards lower energies by $\delta_0$-term of the electron-hole exchange. Some additional irregularities in spacing of the energy levels (visible in Fig.~\ref{fig0}) are due to the spin-flip terms of the $\mathcal{H}_X$ Hamiltonian. First, the anisotropic part of electron-hole exchange mixes the bright states $\left|\pm1,S_z\right\rangle$ associated with the same projection of the Mn$^{2+}$ spin, leading to partial linear polarization of the excitonic transitions \cite{Leger_PRL05}. Second, the off-diagonal terms of the $s,p$-$d$ exchange interaction couple the bright and dark states with the Mn$^{2+}$ spin projection different by~1~\cite{Besombes_PRL04,Leger_PRL05,Leger_PRB07}. This effect is of special importance, since it transfers non-negligible oscillator strength to the dark states allowing for their radiative recombination, which is accompanied by simultaneous flip of the Mn$^{2+}$ spin \cite{Goryca_PRB10}. Nevertheless, the resulting recombination rates of the dark states are typically relatively low, since the amplitude of the spin-flip terms in $\mathcal{H}_X$ is smaller as compared to the amplitude of the Ising terms (especially the isotropic electron-hole exchange and $A_hS_zj_z$ term of the hole-ion exchange). As such, in most of the cases the eigenstates of the $\mathcal{H}_X$ Hamiltonian exhibit either mostly bright or mostly dark character. More specifically, each of these eigenstates is dominated by a single spin state, which allows us to label the eigenstates as $|\psi^X_{J_z,S_z}\rangle$, where $|J_z,S_z\rangle$ corresponds to the dominant admixture to $|\psi^X_{J_z,S_z}\rangle$.

For the full description of the system we also need the Hamiltonian of a biexciton interacting with the Mn$^{2+}$ ion, and the Hamiltonian of the Mn$^{2+}$ spin in an empty QD. These two groups of states have a similar nature, since the 2X is a spin-singlet and thus it is not affected by any exchange interaction. As a result, both the six spin states of the Mn$^{2+}$ ion and the six states of the 2X-Mn$^{2+}$ system are almost degenerate. However, for our considerations it is crucial to note the existence of a small splitting between the states among each group, which in the leading order is given by an effective Hamiltonian of $DS_z^2$ with $D<0$. In the case of the Mn$^{2+}$ ion in the empty dot, this term is related to a residual interaction between the ion and a strained semiconductor lattice \cite{LeGall_PRL09,Goryca_PRL_2014}, while for the 2X it is due to a perturbation of the hole wave function by the hole-Mn$^{2+}$ exchange interaction \cite{Besombes_PRB05,Trojnar_PRL_2011,Trojnar_PRB_2013}. In both cases the value of $|D|$ is of the order of 10~$\mu$eV \cite{LeGall_PRL09,Goryca_PRL_2014,Besombes_PRB_2014,Smolenski_PRB_2015_x2m}, thus being much smaller than any other exchange constant relevant for the X-M$^{2+}$ system. As such, we will neglect the resulting splitting of the Mn$^{2+}$ and 2X-Mn$^{2+}$ eigenstates in the further analysis. Nevertheless, even small $DS_z^2$ term ensures that these eigenstates can be labeled by $S_{z}$ projections of the Mn$^{2+}$ spin onto the QD growth axis $z$ as $|\emptyset,S_z\rangle$ for the Mn$^{2+}$ and $|2\mathrm{X},S_z\rangle$ for the 2X-Mn$^{2+}$ system (here we neglected the hyperfine coupling between the Mn$^{2+}$ electronic and nuclear spins, which is much smaller even compared to $DS_z^2$ term).

\subsection{Modeling of the time-evolution of the system}

The spin dynamics of the Mn$^{2+}$ ion in a quasi-resonantly excited QD is described in terms of the $36$ states of the above-discussed Hamiltonians: $6$ states $|\emptyset, S_z\rangle$ of the empty dot, $24$ eigenstates $|\psi^X_{J_z,S_z}\rangle$ of the $\mathcal{H}_X$ Hamiltonian, and $6$ biexciton states $|2\mathrm{X}, S_z\rangle$. In all the calculations we will assume that the time-evolution of carriers, their spins, and the Mn$^{2+}$ spin, is incoherent -- the state of the whole system is going to be described simply by time-dependent probabilities $p(|\varphi\rangle,t)$ of finding it in one of the aforementioned eigenstates $|\varphi\rangle$. Furthermore, the dynamics will be described by rate equations, with transition rates (due to interaction with photons and phonons) given by Fermi Golden Rule expressions. We thus exclude (rather special) cases in which the characteristic timescale for transitions from one state to two (or more) states is shorter than the inverse of energy bandwidth of the final states, leading to possibility of spontaneous generation of coherence.

\section{Mn$^{2+}$ spin orientation due to appearance of spin-polarized bright excitons in the dot} 
\label{sec:minimal_model}
First we consider the simplest model in which we assume that the state of exciton created in the light absorbing dot does not undergo any change (spin relaxation) during the transfer between the dots, and during its energy relaxation towards the lowest-energy radiative state in the Mn-containing dot. Spin relaxation of carriers and Mn$^{2+}$ due to their coupling to phonons is also excluded. In this case, when the absorbing dot is excited with $\sigma^\pm$ polarized light, the excitons that enter the light-emitting dot have $J_z$ equal to $\pm 1$.

\begin{figure*}
\includegraphics{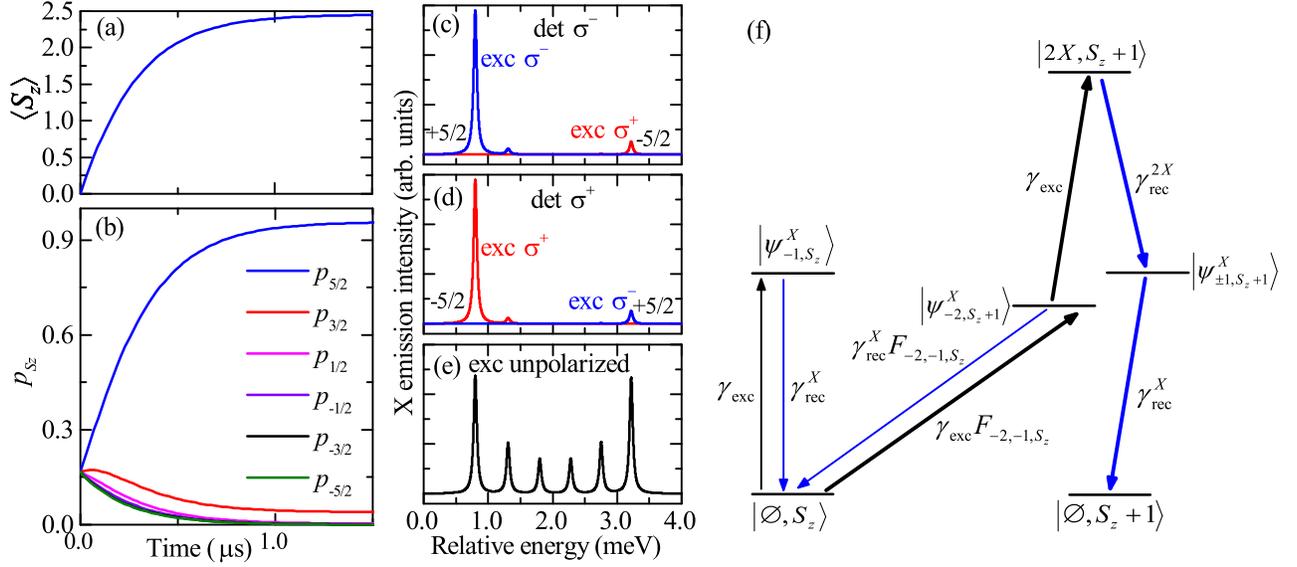}
\caption{(a), (b) Time dependencies of the ion mean spin $\langle S_z\rangle$ and the probabilities $p_{S_z}$ of the occupation of different ion spin states determined upon $\sigma^-$-polarized optical pumping with $1/\gamma_\mathrm{exc}=2$~ns $A_h=0.8$~meV, and for the following QD parameters: $A_e=0.16$~meV, $\delta_0=1.0$~meV, $\delta_1=0$, $1/\gamma^\mathrm{X}_{\mathrm{rec}}=400$~ps, $1/\gamma^\mathrm{2X}_{\mathrm{rec}}=250$~ps. (c), (d) Steady-state X PL spectra computed under circularly polarized excitation for two different helicities of the polarization of detection: $\sigma^-$, $\sigma^+$. (e) Steady-state X PL spectrum due to unpolarized excitation. (f) The scheme presenting the part of state-space, which illustrates the mechanism responsible for the Mn$^{2+}$ spin orientation under $\sigma^-$-polarized excitation in the absence of hh-lh mixing (i.e., for $\epsilon=0$). Bolded arrows indicate the sequence of transitions between the states that leads to increase of the ion spin projection $S_z$ by 1.\label{fig1}}
\end{figure*}

Focusing on  $\sigma^-$ excitation, for example, we have the following possible transitions to consider in the rate equations:
\begin{itemize}
\item exciton creation --- $\gamma(|\emptyset, S_z\rangle\rightarrow|\psi^X_i\rangle)=\gamma_{\mathrm{exc}}|\langle \psi^X_i|\mathcal{P}^\dagger_{-1}|\emptyset, S_z\rangle|^2$,
\item biexciton creation --- $\gamma(|\psi^X_i\rangle\rightarrow|2\mathrm{X}, S_z\rangle)=\gamma_{\mathrm{exc}}\sum_{J_z}|\langle 2\mathrm{X}, S_z|\mathcal{P}^\dagger_{J_z}|\psi^X_i\rangle|^2$,
\item exciton recombination --- $\gamma(|\psi^X_i\rangle\xrightarrow{\sigma^{\pm}}|\emptyset, S_z\rangle)=\gamma^\mathrm{X}_{\mathrm{rec}}|\langle \emptyset, S_z|\mathcal{P}_{\pm1}|\psi^X_i\rangle|^2$ for emission of $\sigma^\pm$ polarized light,
\item biexciton recombination --- $\gamma(|2\mathrm{X}, S_z\rangle\xrightarrow{\sigma^{\pm}}|\psi^X_i\rangle)=\frac{1}{2}\gamma^\mathrm{2X}_{\mathrm{rec}}|\langle\psi^X_i|\mathcal{P}_{\pm1}|2\mathrm{X}, S_z\rangle|^2$ for emission of $\sigma^\pm$ polarized light,
\end{itemize}
where $\mathcal{P}_{J_z}|2\mathrm{X},S_z\rangle=\left|-J_z,S_z\right\rangle$, $\gamma(|\varphi_1\rangle\rightarrow|\varphi_2\rangle)$ is the rate of transition between $|\varphi_1\rangle$ state to $|\varphi_2\rangle$ state, $\gamma_\mathrm{exc}$ is the rate with which the excitons are captured in the orbital ground state of the light-emitting dot (note that this quantity is directly proportional to the power of light exciting the absorbing dot), and $\gamma^\mathrm{X}_{\mathrm{rec}}$ and $\gamma^\mathrm{2X}_{\mathrm{rec}}$ are the inverse lifetimes of bright exciton and biexciton, respectively. For the former one we assume that the intrinsic radiative lifetime of dark exciton that is controlled by hh-lh mixing in CdTe QDs \cite{Smolenski_PRB12} is very long, and the recombination of mostly dark states considered here is due Mn-induced mixing of dark and bright X states. For excitation power much smaller than the one corresponding to saturation, we have $\gamma_\mathrm{exc}\ll\gamma^\mathrm{X}_{\mathrm{rec}}<\gamma^\mathrm{2X}_{\mathrm{rec}}$.

Each state $|\varphi\rangle$ in the model is assigned with its time-dependent occupation probability $p(|\varphi\rangle, t)$. The temporal evolution of these probabilities is computed numerically under an assumption that initially (i.e., at $t=0$) the light-emitting QD is empty, and that the Mn$^{2+}$ spin projection is random (i.e., $p(|\emptyset, S_z\rangle, 0)=1/6$ for any possible $S_z$). The change of the ion spin orientation induced by the injection of spin-polarized excitons to the QD is then quantitatively described by probabilities $p_{S_z}(t)$ of finding the ion in a given spin state $S_z$. These probabilities are obtained by tracing out over the excitonic part of the total wave function:
\begin{eqnarray}
p_{S_z}(t)=&&p(|\emptyset, S_z\rangle, t)\\\nonumber
&+&\sum_i p(|\psi^X_i\rangle, t)\sum_{J_z}|\langle \emptyset, S_z|\mathcal{P}_{J_z}|\psi^X_i\rangle|^2\\\nonumber
&+&p(|2\mathrm{X}, S_z\rangle, t).
\end{eqnarray}
Finally, we also introduce the time-dependent mean spin of the Mn$^{2+}$ ion $\langle S_z\rangle(t)=\sum_{S_z=-5/2}^{5/2}p_{S_z}(t)S_z$. We note here that this quantity does not necessarily correspond to the ion mean spin being typically = obtained in the experiment based on weighted average of the PL intensities of six different emission lines originating from the bright X recombination in Mn-doped QD~\cite{Goryca_PRL09,Goryca_PRB_2015}.

Within this model the orientation of Mn$^{2+}$ occurs due to exchange-induced mixing of bright and dark X states corresponding to distinct $S_z$ projections of the Mn$^{2+}$ spin \cite{Goryca_PRB10}. Such mixing is caused by flip-flop parts of exchange interaction. These are always present for the isotropic e-Mn$^{2+}$ exchange, but for h-Mn$^{2+}$ exchange the mixing of hh and lh states, leading to finite $\epsilon$ in Eq.~(\ref{eq:Hspd}), is necessary for their existence. These two interactions, when treated separately, lead to Mn$^{2+}$ orientation dynamics of qualitatively distinct character. Below we first consider them one-by-one, and then we give results for the realistic case in which both are contributing to the dynamics.

\subsection{Orientation due to electron-Mn$^{2+}$ exchange interaction only}
\label{subsec:onlyeMn}

With $\epsilon=0$ in $\mathcal{H}_X$ Hamiltonian only the electron spin can flip-flop with the Mn$^{2+}$ spin. In such a case the above model leads to the orientation of Mn$^{2+}$ spin towards the $\pm5/2$ state when excitons of $J_z=\mp1$ are created in the Mn-containing dot. Time dependencies of the average Mn$^{2+}$ spin $\langle S_z\rangle$ and Mn$^{2+}$ spin level occupations $p_{S_z}$ for $\sigma^{-}$ excitation are shown in Fig.~\ref{fig1}(a) and \ref{fig1}(b), while the steady-state PL spectra of the neutral exciton (due to polarized or unpolarized excitations) are shown in Figs.~\ref{fig1}(c)-\ref{fig1}(e).

\begin{figure}
\centering
\includegraphics{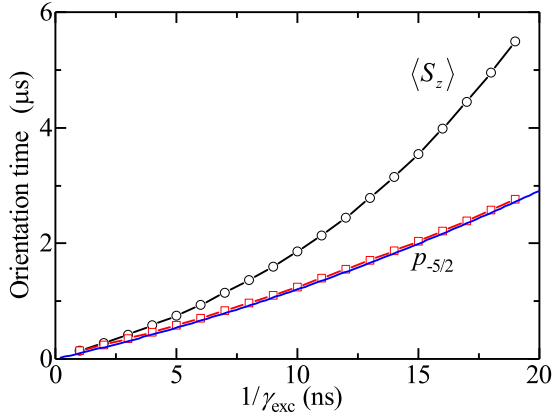}
\caption{(a) The characteristic orientation times of the Mn$^{2+}$ spin plotted as a function of the inverse (relative) excitation power $1/\gamma_\mathrm{exc}$ of circularly-polarized optical excitation. Both presented times were determined from the exponential fit of numerically-calculated dependencies of $\langle S_z\rangle(t)$ (circles) and $p_{-5/2}(t)$ (squares) (the calculations were carried out for the QD parameters listed in the caption of Fig.~\ref{fig1}). The solid line represents the characteristic time of the Mn$^{2+}$ spin-flip from $S_z=-5/2$ to $S_z=-3/2$, obtained from the simplified analytical formula given by Eq. (\ref{Eq:eMnsimplerate}).\label{fig2}}
\end{figure}

The physical mechanism of Mn$^{2+}$ spin orientation can be understood most easily by looking at a subset of states and at transitions between them shown in Fig.~\ref{fig1}(f). We assume that initially there is no exciton in the light-emitting dot, and that the magnetic ion has spin projection $S_z<5/2$. Under $\sigma^-$ excitation the exciton with $J_z=-1$ appears with the largest probability in mostly bright state $|\psi^X_{-1,S_z}\rangle$ (the subscript denotes the pure spin state corresponding to a dominant admixture in the considered eigenstate, see Sec.~\ref{subseq:Model_Hamiltonian}). In this case the exciton subsequently undergoes radiative recombination, the system returns to $|\emptyset, S_z\rangle$, and the ion spin remains unchanged. However, there is a nonzero probability that the incoming $J_z=-1$ exciton falls into the mostly dark state $|\psi^X_{-2,S_z+1}\rangle$, which contains an admixture of $|-1,S_z\rangle$ due to $A_eS_+\sigma_-/2$ flip-flop part of the e-Mn$^{2+}$ exchange interaction. The probability corresponding to this admixture is $F_{-2,-1,S_z} \! \equiv\! |\langle-1,S_z|\psi^X_{-2,S_z+1}\rangle|^2$. The radiative recombination of this exciton is thus possible, but its rate $\gamma^\mathrm{X}_{\mathrm{rec}}F_{-2,-1,S_z}$ is typically (i.e.~for typical excitation power) much smaller than the rate $\gamma_\mathrm{exc}$ of addition of a second exciton to the dot, i.e.~of $|\mathrm{2X}, S_z+1\rangle$ biexciton creation. Subsequent cascaded recombination of this complex leaves the system in $|\emptyset, S_z+1\rangle$ state, leading to change of ion spin by $1$. The rate of this process can be approximately written as
\begin{equation}
\gamma_e(S_z\xrightarrow{\sigma^{-}} S_z+1)\simeq\gamma_{\mathrm{exc}}F_{-2,-1,S_z}\frac{\gamma_{\mathrm{exc}}}{\gamma_{\mathrm{exc}}+\gamma^\mathrm{X}_{\mathrm{rec}}F_{-2,-1,S_z}}.
\label{Eq:eMnsimplerate}
\end{equation}
In the simple model considered now there are no comparably effective processes leading to opposite direction of Mn$^{2+}$ spin orientation (i.e.~$\gamma_e(S_z+1\xrightarrow{\sigma^{-}} S_z)\simeq0$), so that the above formula gives the rate for optical orientation of Mn$^{2+}$ spin.

Note that for excitation power fulfilling $\gamma_{\mathrm{exc}} \gg \gamma^\mathrm{X}_{\mathrm{rec}}F_{-2,-1,S_z}$ (which is easily fulfilled while keeping the power below the saturatiom value), the orientation rate is approximately equal to $\gamma_{\mathrm{exc}}F_{-2,-1,S_z}$, i.e., to the probability of creation of the first exciton in the mostly dark state. The orientation rate in this regime is thus linearly proportional to the excitation power. 

The above analytical quantitative result is corroborated by numerical simulation of the dynamics of the whole system. In Figs. \ref{fig1}(a) and \ref{fig1}(b) we show the time dependencies of $\langle S_z\rangle(t)$ and $p_{S_z}(t)$. Both of them can be fit with an exponential function parametrized by an orientation time. The excitation power dependencies of these times are shown in Fig.~\ref{fig2}, where we also show the agreement between the fitted orientation time for $p_{-5/2}(t)$ and the time obtained by considering only the $-5/2$ to $-3/2$ transition with the rate from Eq.~(\ref{Eq:eMnsimplerate}). We note also that for lower excitation powers (i.e., larger values of $1/\gamma_\mathrm{exc}$) this time becomes clearly shorter than the orientation time determined based on the time dependence of the ion mean spin $\langle S_z\rangle(t)$. This effect can be qualitatively understood by taking into account that, in contrast to $p_{-5/2}(t)$, the change of $\langle S_z\rangle$ involves depletion and filling of subsequent Mn$^{2+}$ spin states, thus being a slower process as compared to depletion of $S_z=-5/2$ state. This is also the reason for the difference in the range in which both studied orientation times exhibit linear excitation power dependence (see Fig.~\ref{fig2}).

Finally, let us turn our attention to the predicted shape of PL signal for unpolarized excitation shown in Fig.~\ref{fig1}(e). The mean value of the Mn$^{2+}$ ions spin is of course zero then, but the intensities of lines corresponding to distinct values of $|S_{z}|$ are unequal, and the strongest lines in the spectrum correspond to $|S_z| = 5/2$. Such an obviously non-equilibrium pattern of $p_{S_z}$ occupations is caused by dependence of rates of the electron-induced Mn$^{2+}$ spin-flip on the initial state of the ion spin. For $\sigma^+$ excitation the flip rate at high excitation power is $\gamma_e(S_z+1\xrightarrow{\sigma^{+}}S_z)\simeq\gamma_{\mathrm{exc}}F_{2,1,S_z+1}$, where $F_{2,1,S_z+1}=|\langle+1,S_z+1|\psi^X_{+2,S_z}\rangle|^2$. On the other hand, for $\sigma_{-}$ excitation, the $\gamma_e(S_z\xrightarrow{\sigma^{-}}S_z+1)$ is determined by $F_{-2,-1,S_z}$ coefficient. In the lowest order of perturbation theory, while neglecting $\delta_1$ splitting, we have
\begin{eqnarray}
F_{-2,-1,S_z}&=&\left|\frac{A_e\sqrt{S(S+1)-S_z(S_z+1)}}{2\delta_0+A_h-A_e(2S_z+1)}\right|^2,\\\nonumber F_{2,1,S_z+1}&=&\left|\frac{A_e\sqrt{S(S+1)-S_z(S_z+1)}}{2\delta_0+A_h+A_e(2S_z+1)}\right|^2,
\end{eqnarray}
where $S=5/2$ for Mn$^{2+}$. The resulting ratio of the transition probabilities is then given (at large $\gamma_\mathrm{exc}$) by
\begin{multline}
\frac{\gamma_e(S_z\xrightarrow{\sigma^{-}}S_z+1)}{\gamma_e(S_z+1\xrightarrow{\sigma^{+}}S_z)}\approx\\\approx\frac{F_{-2,-1,S_z}}{F_{2,1,S_z+1}}=\left(\frac{2\delta_0+A_h+A_e(2S_z+1)}{2\delta_0+A_h-A_e(2S_z+1)}\right)^2 \,\, .
\label{eq:diabelek}
\end{multline}
This ratio is larger (smaller) than $1$ for $S_z>0$ ($S_z<0$). Consequently, during the excitation with unpolarized light, when excitons with random spin $J_{z} = \pm 1$ are created, the two processes of driving the Mn$^{2+}$ spin to $S_{z} =\pm 5/2$ state are coexisting, and the net result is a characteristic spectrum shape from Fig.~\ref{fig1}(e).

\subsection{Orientation due to hole-Mn$^{2+}$ exchange interaction only}
\label{subsec:onlyhMn}

\begin{figure*}[t]
\includegraphics{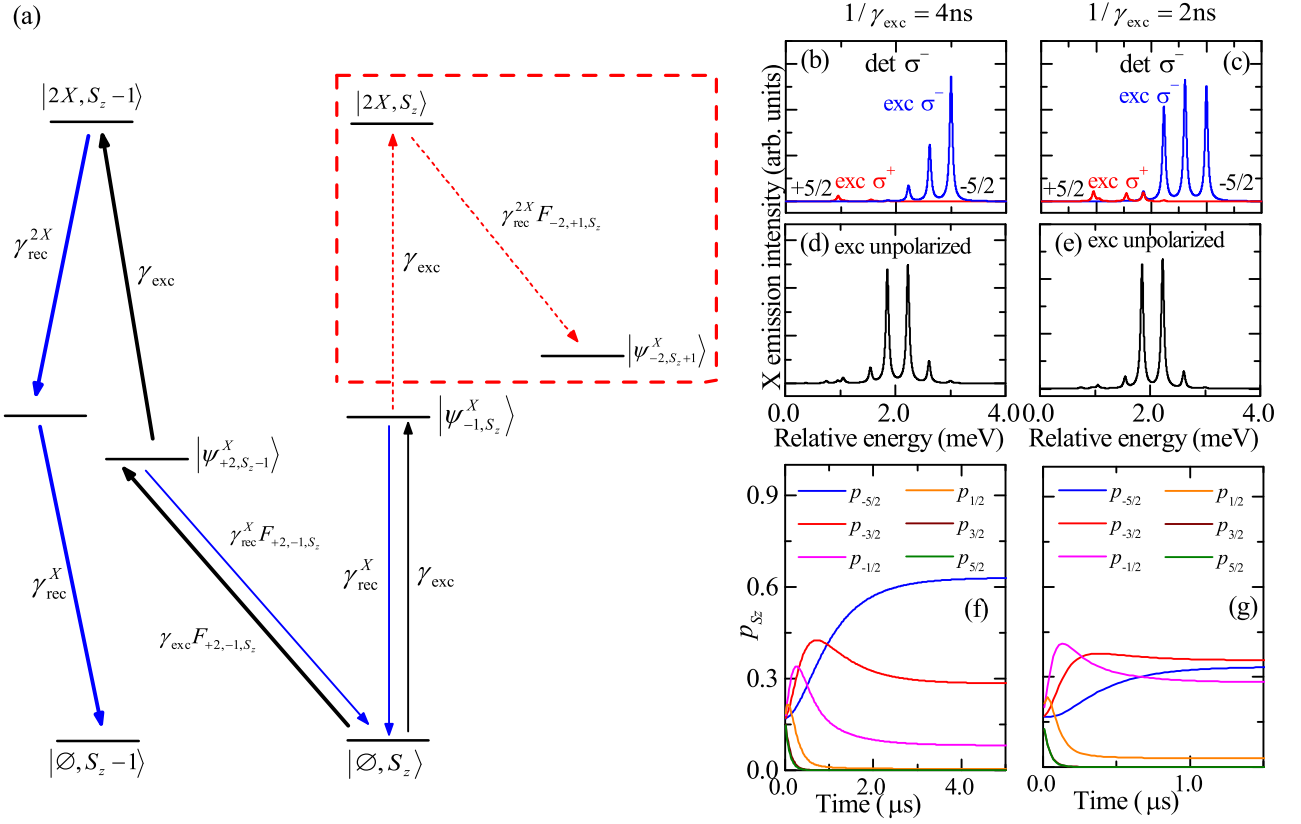}
\caption{(a) The scheme illustrating the mechanism responsible for the Mn$^{2+}$ spin orientation under $\sigma^-$-polarized excitation in the presence of non-zero hh-lh mixing $\epsilon\neq0$, but for $A_e=0$. Bolded arrows indicate the sequence of transitions between the states that leads to decrease of the Mn$^{2+}$ spin projection $S_z$ by 1. Dashed arrows (as well as the part of states plotted inside the dashed rectangle) exemplify the process that may increase $S_z$ by 1 for sufficiently large excitation powers. (b-e) Steady-state X PL spectra computed for: (b),(c) circularly-polarized excitation, (d),(e) unpolarized excitation. All spectra correspond to $\sigma^-$ polarization of detection. The spectra in (b) and (d) were calculated for $1/\gamma_\mathrm{exc}=4$~ns, while the spectra in (c) and (e) for $1/\gamma_\mathrm{exc}=2$~ns. (f), (g) Time dependencies of $p_{S_z}$ determined for these excitation powers for the following QD parameters: $A_h=0.8$~meV, $\epsilon=0.1$, $A_e=0$, $\delta_0=1.0$~meV, $\delta_1=0$, $1/\gamma^\mathrm{X}_{\mathrm{rec}}=400$~ps, and $1/\gamma^\mathrm{2X}_{\mathrm{rec}}=250$~ps.
\label{fig3}}
\end{figure*}

Let us neglect now the electron-Mn$^{2+}$ exchange, but assume finite hh-lh mixing leading to possibility of hole-Mn$^{2+}$ flip-flop. A mechanism analogous to the one described above for electron-Mn$^{2+}$ flip-flop lead then to optical orientation of Mn$^{2+}$. However, due to the fact that the spins of the electron and the hole forming the bright exciton are antiparallel, the direction of the resulting orientation has an opposite direction to the one obtained before: creation of $J_z=\pm1$ excitons increases the occupation of $S_z=\pm5/2$ levels. In Fig.~\ref{fig3}(a) we show the scheme of transitions leading to the Mn$^{2+}$ spin orientation. The mechanism is the same as before. For example, if we start from occupied $S_z>-5/2$ state, $J_z=-1$ exciton created by $\sigma^-$ light can, upon entering the emitting dot, find itself in a mostly dark $|\psi^X_{2,S_z-1}\rangle$  state with rate $\gamma_{\mathrm{exc}}F_{2,-1,S_z}$ (where $F_{2,-1,S_z}=|\langle-1,S_z|\psi^X_{2,S_z-1}\rangle|^2$). Subsequent creation of biexciton and its two-step recombination leads to diminishing of Mn$^{2+}$ spin by 1. The rate for the latter process is given by a formula analogous to the one from Eq.~(\ref{Eq:eMnsimplerate}):
\begin{equation}
\gamma_h(S_z\xrightarrow{\sigma^{-}} S_z-1)\simeq\gamma_{\mathrm{exc}}F_{2,-1,S_z}\frac{\gamma_{\mathrm{exc}}}{\gamma_{\mathrm{exc}}+\gamma^\mathrm{X}_{\mathrm{rec}}F_{2,-1,S_z}}.
\label{Eq:hMnsimplerate}
\end{equation}

It is important to note that in this case the $F_{\pm 2,\mp 1,S_z}$ parameters that determine the orientation dynamics, exhibit much larger variation as a function of $S_z$ than the $F_{\pm 2,\pm 1,S_z}$ coefficients that were determining the electron-related dynamics. This is caused by the fact that the hole-Mn$^{2+}$ exchange $A_{h}$ is significant compared to all the other energies relevant for the complex of Mn$^{2+}$ and the exciton in its orbital ground state.  $F_{2,-1,S_z}$ describes mixing between bright and dark states differing by the hole spin flip, and the energy differences between these states show much larger variation than the energy differences between X states differing by the electron spin flip. In the lowest order of perturbation theory, neglecting again the small corrections due to $\delta_1$, we have
\begin{equation}
F_{\pm2,\mp1,S_z}=\left|\frac{2A_h\epsilon\sqrt{S(S+1)-S_z(S_z\mp1)}}{2\delta_0\mp A_h(2S_z\mp1)-A_e}\right|^2.
\label{eq:F2m1Sz} 
\end{equation}
For the typical values of $A_{h} \! \approx \! \delta_{0} \! \approx \! 1$ meV, the denominator of this expression can be either rather small (a fraction of a meV) for $S_z$ close to $5/2$, while for $S_z = -5/2$ its value can reach a few meV. 

The visibly different character of $p_{S_z}(t)$ time dependence for various $S_z$, shown in Figs.~\ref{fig3}(f) and \ref{fig3}(g) for $\sigma_{-}$ excitation, is the first consequence of the above observation. While in Fig.~{fig1}(b) the populations undergoing depletion were all decaying on the same timescale, in Figs.~\ref{fig3}(f) and \ref{fig3}(g) we see that now the occupations of $S_{z} > 0$ levels decay much faster than the others.

A characteristic PL spectrum shape predicted for unpolarized excitation is shown in Fig.~\ref{fig3}(d). As in the previous Section we see unequal intensities of lines corresponding to distinct $|S_z|$, but the intensity ratios are much higher, and the pattern is completely different: now, instead of six peaks commonly used for identification of a QD containing a single Mn$^{2+}$ spin, we see only two peaks corresponding to $S_z=\pm1/2$. This can be explained in a way analogous to the one presented in the previous Section. Let us just reproduce the result for the ratio of transition rates:
\begin{multline}
\frac{\gamma_h(S_z\xrightarrow{\sigma^{-}}S_z-1)}{\gamma_h(S_z-1\xrightarrow{\sigma^{+}}S_z)}\approx\\\approx\frac{F_{2,-1,S_z}}{F_{-2,1,S_z-1}}=\left(\frac{2\delta_0+A_h(2S_z-1)-A_e}{2\delta_0-A_h(2S_z-1)-A_e}\right)^2 \,\, ,
\end{multline}
which gives, for typical parameters, the values larger (smaller) than 1 for $S_z>0$ ($S_z<0$). Significant range of values of $F_{2,-1,S_z}/F_{-2,1,S_z-1}$ ratios for various $S_{z}$, resulting in large differences of the intensities between the lines in the spectrum, is the second consequence of strong sensitivity of $F_{\pm 2,\mp 1,S_z}$ to the $S_{z}$ value. 

Finally, let us notice an interesting behavior that occurs as we increase the excitation intensity from $\gamma_\mathrm{exc}=(4~\mathrm{ns})^{-1}$ to $\gamma_\mathrm{exc}=(2~\mathrm{ns})^{-1}$, compare the left and right panels of Fig.~\ref{fig3}. Interestingly, for larger power the Mn$^{2+}$ spin level most occupied in steady state is $S_z=-3/2$, not $-5/2$. This can also be explained by strong $S_z$ dependence of bright-dark state mixing. If a biexciton in $|2\mathrm{X},S_z\rangle$ state gets created (which is an event which occurs more often at larger power), it recombines into a mostly bright $|\psi^X_{\pm1,S_z}\rangle$ state with $\gamma^{\mathrm{2X}}_\mathrm{rec}/2$ rate and into a mostly dark $|\psi^X_{\pm2,S_z\mp1}\rangle$ state with  $\gamma^{\mathrm{2X}}_\mathrm{rec}F_{\pm2,\mp1,S_z}/2$ rate. The latter is much smaller than the former, but for parameters used in  \ref{fig3} we have  $F_{-2,1,-3/2}$ mixing coefficient more than 200 times larger than $F_{2,-1,-3/2}$ due to small energy difference between  $|\psi^X_{1,-3/2}\rangle$ and $|\psi^X_{-2,-1/2}\rangle$ states. As a consequence, the recombination of the biexciton from $|2\mathrm{X},-3/2\rangle$ to $|\psi^X_{-2,-1/2}\rangle$ is an effective channel of Mn$^{2+}$ flipping from $-3/2$ do $-1/2$ state. This is schematically shown in \ref{fig3}(a). Note, however, that the presence of this effect depends sensitively on the dot parameters.

\subsection{Orientation due to both electron- and hole-Mn$^{2+}$ exchange}
\begin{figure}
\centering
\includegraphics{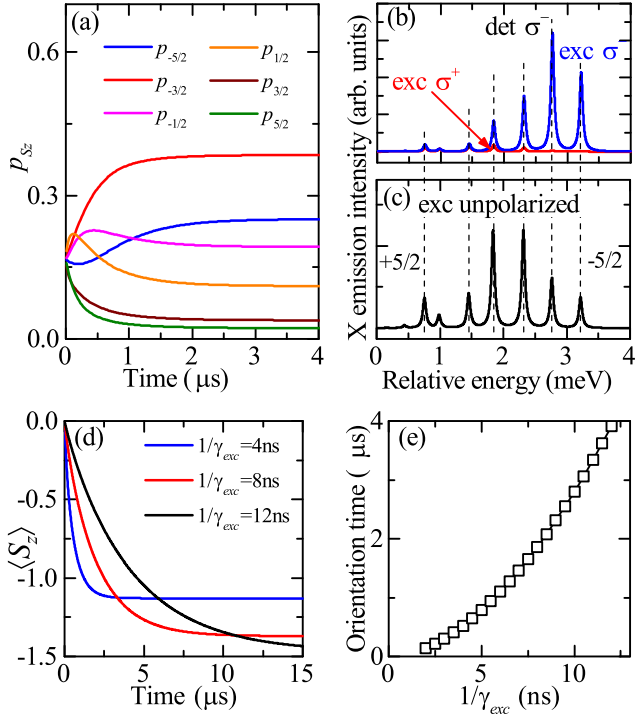}
\caption{(a) Time dependencies of $p_{S_z}$ due to $\sigma^-$-polarized excitation computed numerically for $1/\gamma_\mathrm{exc}=4$~ns, $A_h=0.8$~meV, $A_e=0.16$~meV, $\epsilon=0.1$, $\delta_0=1.0$~meV, $\delta_1=0$, $1/\gamma^\mathrm{X}_{\mathrm{rec}}=400$~ps, $1/\gamma^\mathrm{2X}_{\mathrm{rec}}=250$~ps. (b), (c) Steady-state X PL spectra under $\sigma^-$-polarized and unpolarized excitation. (d) Time dependencies of $\langle S_z\rangle$ computed for the same set of QD parameters and various intensities of $\sigma^-$-polarized excitation. (e) The orientation time determined by fitting $\langle S_z\rangle(t)$ with an exponential function versus the inverse (relative) excitation power $1/\gamma_\mathrm{exc}$.\label{fig4}}
\end{figure}

\begin{figure*}[t!]
\includegraphics{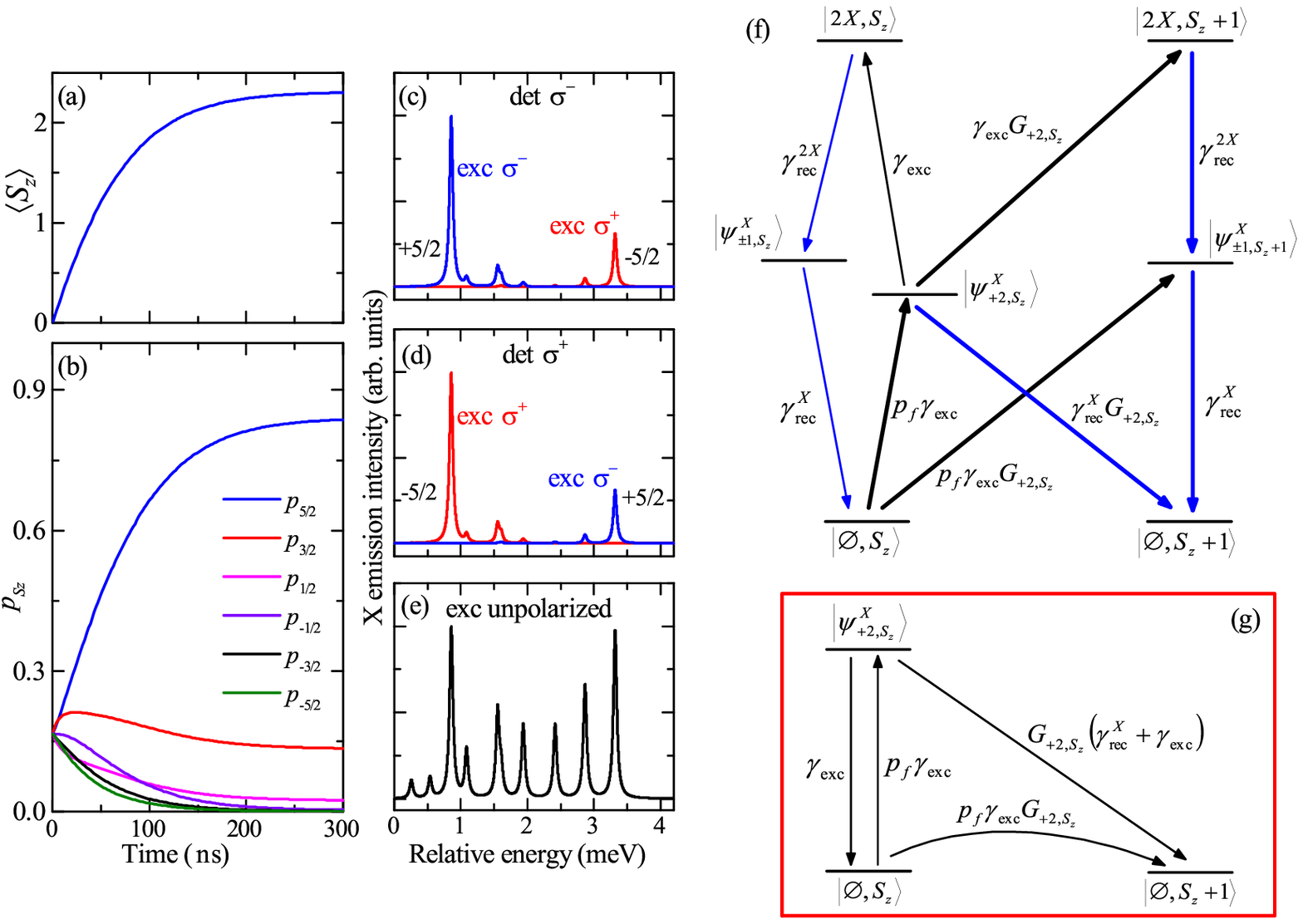}
\caption{(a), (b) Numerically computed time dependencies of $\langle S_z\rangle$ and $p_{S_z}$ for $\sigma^-$-polarized excitation under an assumption that the probability of the hole spin-flip during the inter-dot transfer is equal to $p_f=40$\%. All other model parameters were taken to be the same as for the simulation presented in Fig.~\ref{fig4}, i.e., $1/\gamma_\mathrm{exc}=4$~ns, $A_h=0.8$~meV, $A_e=0.16$~meV, $\epsilon=0.1$, $\delta_0=1.0$~meV, $\delta_1=0$, $1/\gamma^\mathrm{X}_{\mathrm{rec}}=400$~ps, $1/\gamma^\mathrm{2X}_{\mathrm{rec}}=250$~ps. (c), (d) Steady state X PL spectra under circularly-polarized excitation for two different circular polarizations of detection: $\sigma^-$, $\sigma^+$. (e) Steady-state X PL spectrum under unpolarized excitation (f) The scheme illustrating the mechanism responsible for the Mn$^{2+}$ spin orientation due to injection of dark excitons (created by the absorption of a $\sigma^-$-polarized photon and subsequent spin-flip of the hole). Bolded arrows represent the sequence of inter-state transitions that leads to increase of the Mn$^{2+}$ spin projection by 1. (g) Simplified diagram of states (relevant for $\gamma_\mathrm{exc}\ll\gamma^\mathrm{X}_{\mathrm{rec}}<\gamma^\mathrm{2X}_{\mathrm{rec}}$), which allows to analytical determination of approximate rate of the Mn$^{2+}$.\label{fig5}}
\end{figure*}

Since e-Mn$^{2+}$ and h-Mn$^{2+}$ interactions lead to Mn$^{2+}$ spin orientation in opposite directions, the steady state occupations of $S_{z}$ levels in the realistic case of coexistence of the two interactions will depend on their relative strength. In fact, using various sets of experimentally determined parameters of CdTe/ZnTe QDs, orientation in either direction (i.e.~towards dominant $S_z=\pm5/2$ population) can be theoretically obtained. In Fig.~\ref{fig4} we show results for parameters used previously in Fig.~\ref{fig1}, only with nonzero value of valence subband mixing  ($\epsilon=0.1$). The orientation due to h-Mn$^{2+}$ exchange is then dominant, but the presence of e-Mn$^{2+}$ interaction leads to finite population of  $S_z>0$ states under $\sigma^-$ illumination (cf.~Fig.~\ref{fig3}), and the intensities of PL lines under linear polarization excitation in Fig.~\ref{fig4}(c) are less inhomogeneous than in Fig.~\ref{fig3}(d). The power dependence of Mn$^{2+}$ spin dynamics shown in Fig.~\ref{fig4}(e) is very similar to the one obtained in $\epsilon \! =\! 0$ case, cf.~Fig.~\ref{fig2}(a). This follows from identical dependence of $\gamma_e$ and $\gamma_h$ on $\gamma_\mathrm{exc}$, see Eqs.~(\ref{Eq:eMnsimplerate}) and (\ref{Eq:hMnsimplerate}). For any choice of model parameters the characteristic timescale of change of $S_{z}$ occupation is  $\tau_\mathrm{orient}\propto\tau_\mathrm{exc}(1+\tau_\mathrm{exc}/\tau_\mathrm{dark})$, where $\tau_\mathrm{exc}=1/\gamma_\mathrm{exc}$ and $\tau_\mathrm{dark}$ is an {\it effective} lifetime of mostly dark states corresponding to given $S_z$.

\section{Influence of exciton spin relaxation on the Mn$^{2+}$ orientation}
\label{sec:dark_and_bright}

\begin{figure*}
\includegraphics{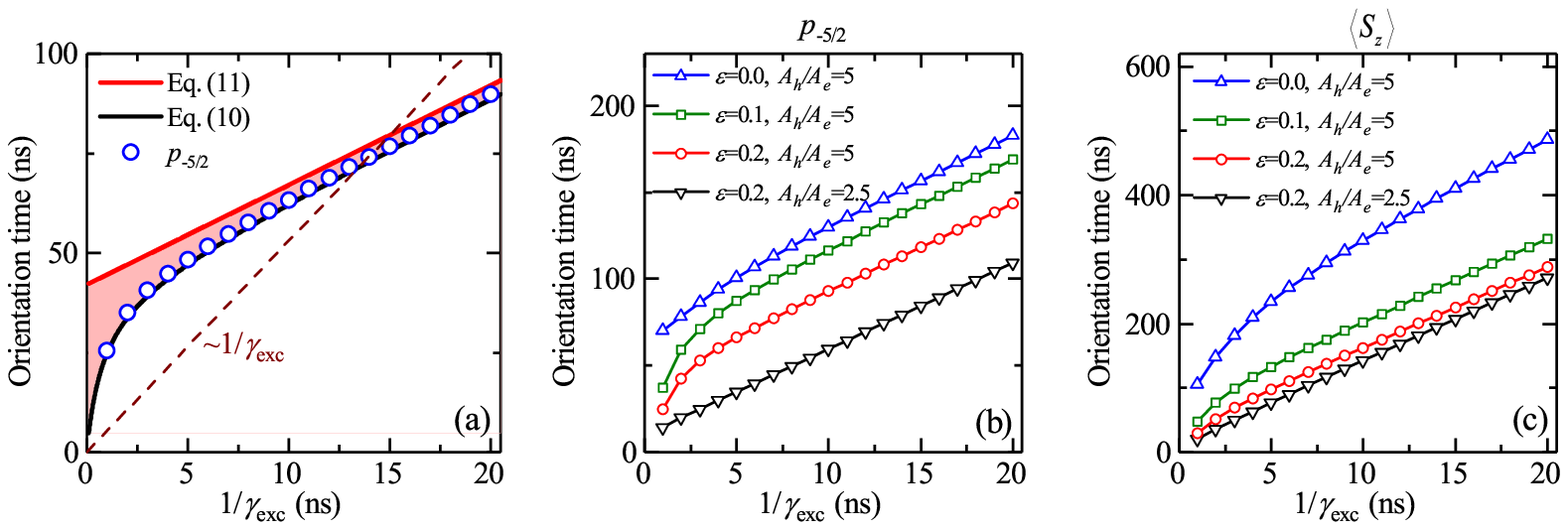}
\caption{(a) Orientation time of the Mn$^{2+}$ spin as a function of the inverse (relative) power $1/\gamma_\mathrm{exc}$ of $\sigma^-$-polarized excitation. The plotted times were determined from the exponential fit of numerically-computed $p_{-5/2}(t)$. The calculations were carried out for $p_f=40$\% and the QD parameters listed in the caption of~\ref{fig5}. Solid lines represent the characteristic time of the Mn$^{2+}$ spin-flip from $S_z=-5/2$ to $S_z=-3/2$ state, obtained from the simplified analytical formulas given by Eq. (\ref{Eq:dMnrate}) and Eq. (\ref{Eq:dMnsimplerate}). The dashed line corresponds to a linear dependence $\tau_\mathrm{orient}\propto1/\gamma_\mathrm{exc}$ known from the experiment. (b),(c) The dependence of the orientation time for (b) $p_{-5/2}(t)$ and (c) $\langle S_z\rangle(t)$ on $1/\gamma_\mathrm{exc}$ determined for different strengths $\epsilon$ of the hh-lh mixing and various ratios of the ion-carrier exchange intergrals $A_h/A_e$ for fixed $A_h=0.8$~meV, $p_f=20$\%, $\delta_0=1.0$~meV, $\delta_1=0$, $1/\gamma^\mathrm{X}_{\mathrm{rec}}=400$~ps, and $1/\gamma^\mathrm{2X}_{\mathrm{rec}}=250$~ps.
\label{fig6}}
\end{figure*}

Let us include now the exciton spin relaxation (occurring without the change of Mn$^{2+}$ spin) in the model. Such a relaxation can occur due to exciton-phonon scattering with the help of either spin-orbit coupling \cite{Khaetskii_PRB01, Woods_PRB02, Hanson_RMP07} or the electron-hole exchange interaction \cite{Tsitsishvili_PRB03, Roszak_PRB07} allowing for the carrier spin-flip.

Spin relaxation of the exciton in the orbital ground states of the QD with the Mn$^{2+}$ spin was invoked to explain experiments on Mn$^{2+}$ spin orientation caused by {\it resonant} excitation of one of X-Mn$^{2+}$ transitions \cite{Cywinski_PRB10,Cao_PRB11}. The relaxation of the hole spin leads to a transition from bright $|J_z=\pm1,S_z\rangle$ state created by $\sigma^\pm$ excitation, to mostly dark $|J_z=\mp2,S_z\rangle$ state. This state is ``brightened'' by the Mn$^{2+}$-carrier exchange interactions, which mix it with the bright states having the dominant component of  $S_z\mp1$. Recombination of the dark state leads then to the orientation of the Mn$^{2+}$ spin towards $Sz = \mp5/2$, in agreement with the experiments \cite{LeGall_PRL09,LeGall_PRB10}.

The above-described simple mechanism requires a finite rate of the spin relaxation of a hole contained in an exciton. This rate is not exactly known for  CdTe/ZnTe QDs, but it can be estimated, e.g., from polarization-resolved measurements of the trion PL \cite{LeGall_PRB_2012}. The latter show that the hole relaxation time is of the order of ten ns. This gives a rough estimate of hole relaxation time $\tau_h$ between X levels split by exchange interaction with Mn$^{2+}$ and by electron-hole exchange. A crucial thing to note is that while for resonant excitation the Mn$^{2+}$ orientation time is comparable to this $\tau_h$, in the quasi-resonant case of interest here the Mn$^{2+}$ orientation rate is much lower than the hole relaxation rate. The reason is simple: under resonant excitation, once a dark state is created by hole spin relaxation, the system stays in this state (as the exciting light becomes nonresonant) until the event of radiative recombination leading to Mn$^{2+}$ spin change. In the quasi-resonant case the introduction of the second exciton to the Mn-containing dot, leading to creation of biexciton, preempts the radiative recombination of the mostly dark state. The subsequent recombination of the biexciton leaves the average Mn$^{2+}$ spin state unchanged. 

Spin relaxation of exciton in the orbital ground state of Mn-containing dot is thus not expected to contribute to the Mn$^{2+}$ spin orientation occurring on timescales of $\lesssim 100$ ns. What remains to be considered is the influence of X spin relaxation {\it during} their transfer from the excited state to the radiatively active orbital ground state. Measurements show that this relaxation is significant: in the case of excitation of the double-dot system about 40\% of bright excitons created resonantly in the Mn-free QD finish up in the Mn-containing dot as mostly dark ones \cite{Smolenski_PRB_2015_2X}. Furthermore, it was proven in the same work that it is the hole spin that becomes almost random during the energy relaxation of the exciton, while the electron spin remains conserved to much larger degree.

In order to take this relaxation into account we modify the model by adding for $\sigma_{\pm}$ excitation:
\begin{itemize}
\item creation of dark exciton with a flipped hole spin --- $\gamma(|\emptyset,S_z\rangle\rightarrow|\psi^X_i\rangle)=p_f\gamma_{\mathrm{exc}}|\langle\psi^X_i|\mathcal{P}_{\mp 2}^\dagger|\emptyset,S_z\rangle|^2$,
\item and diminished probability of bright exciton creation --- $\gamma(|\emptyset,S_z\rangle\rightarrow|\psi^X_i\rangle)=(1-p_f)\gamma_{\mathrm{exc}}|\langle\psi^X_i|\mathcal{P}_{\pm 1}^\dagger|\emptyset,S_z\rangle|^2$,
\end{itemize}
where $p_f$ hole spin-flip probability.

These deceptively minor modifications lead to a qualitative change of the character of Mn$^{2+}$ spin orientation. For large $p_f \sim 40$\% the direction of orientation is always in agreement with observations from \cite{Goryca_PRL09}, independent of the QD parameters. Furthermore, the characteristic orientation timescale becomes about an order of magnitude shorter than in the previously discussed models. This can be seen in Figs.~\ref{fig5}(a),(b) containing results for parameters the same as for Fig.~\ref{fig4}, only with $p_f=40$\% assumed. 

These strong effects can be understood with the help of Fig.~\ref{fig5}(f), in which a scheme of relevant transitions is shown. Let us assume initial Mn$^{2+}$ spin projection $S_z<5/2$ and consider dark $J_z=+2$ exciton (created by  $\sigma^-$ light absorption and subsequent hole spin relaxation) entering the Mn-containing QD (this happens with $p_f\gamma_\mathrm{exc}$ rate). Two things can happen then:
\begin{itemize}
\item with $\gamma^\mathrm{X}_\mathrm{rec}G_{2,S_z}$ rate the exciton will recombine radiatively, leaving the system in  $|\emptyset,S_z+1\rangle$ state with Mn$^{2+}$ spin increased by $1$ (with  $G_{2,S_z}=F_{2,1,S_z+1}+F_{2,-1,S_z+1}=|\langle \psi^X_{+2,S_z}|1,S_z+1\rangle|^2+|\langle \psi^X_{+2,S_z}|{-1},{S_z+1}\rangle|^2$ quantifying the mixing of $|\psi^X_{+2,S_z}\rangle$ state withthe bright states), or
\item another exciton appears and a biexction is created. The most probable state of the biexciton is  $|2\mathrm{X},S_z\rangle$, the recombination of which will leave the Mn$^{2+}$ spin unchanged. Due to a bright state admixture in  $|\psi^X_{+2,S_z}\rangle$ a biexciton in $|2\mathrm{X},S_z+1\rangle$ can be created with $\gamma_\mathrm{exc}G_{2,S_z}$ rate, leading to Mn$^{2+}$ spin increase by $1$ after recombination.
\end{itemize}
Furthermore, incoming $J_z=+2$  exciton can find itself in a bright final state $|\psi^X_{\pm1,S_z+1}\rangle$ with rate $p_f\gamma_\mathrm{exc}G_{2,S_z}$. The subsequent recombination again leads to final $|\emptyset,S_z+1\rangle$ state.

Out of these three sequences of transitions leading to Mn$^{2+}$ spin orientation the most probable one is the dark exciton recombination. This is due to the basic assumption of $\gamma^\mathrm{X}_{rec}\gg\gamma_\mathrm{exc}$. Analytical derivation of the orientation rate is more involved than in the previously discussed models. However, using the fact that the recombination of bright states is the fastest considered process, we can simplify the transition scheme by removing the quickly radiatively depleted states, as shown in Fig.~ \ref{fig5}(g). Using this approximation it is possible to analytically calculate the time-dependence of occupations of $|\emptyset,S_z\rangle$ and $|\psi^X_{+2,S_z}\rangle$ states. These dynamics have multi-exponential character, but we can define an effective timescale $\tau_d$ as the inverse of $\int_0^\infty p_{S_z}(t)dt/P_0$ (where $P_0$ is the initial occupation of  $|\emptyset,S_z\rangle$ state). The rate of $S_z\xrightarrow{\sigma^{-}} S_z+1$ transition can be then approximated by $\gamma_d\! = \! 1/\tau_d$ given by
\begin{equation}
\gamma_d  \approx G_{2,S_z}\gamma_\mathrm{exc}\frac{p_f\left(\gamma^\mathrm{X}_\mathrm{rec}+2\gamma_\mathrm{exc}\right)}{(1+p_f)\gamma_\mathrm{exc}+G_{2,S_z}\gamma^\mathrm{X}_\mathrm{rec}} \,\, .
\label{Eq:dMnrate}
\end{equation}
Since $\gamma^\mathrm{X}_\mathrm{rec}\gg\gamma_\mathrm{exc}$, we have $\gamma_d(S_z\xrightarrow{\sigma^{-}} S_z+1)\gg G_{2,S_z}\gamma_\mathrm{exc}$. Comparing this result with Eqs.~(\ref{Eq:eMnsimplerate}) and (\ref{Eq:hMnsimplerate}) for $\gamma_e$ and $\gamma_h$ we see that indeed $\gamma_d\gg\gamma_e,\gamma_h$. Finally, we can rewrite Eq.~(\ref{Eq:dMnrate}) as
\begin{multline}
\tau_\mathrm{orient}(S_z\xrightarrow{\sigma^{-}} S_z+1)=\frac{1}{\gamma_d(S_z\xrightarrow{\sigma^{-}} S_z+1)}\approx\\ \approx\frac{1+p_f}{p_f}\frac{1}{G_{2,S_z}\gamma^\mathrm{X}_\mathrm{rec}}+\frac{1}{p_f}\frac{1}{\gamma_\mathrm{exc}},
\label{Eq:dMnsimplerate}
\end{multline}
This is an intuitive result: the characteristic Mn$^{2+}$  spin orientation time is a sum of time proportional to dark exciton lifetime, $1/G_{2,S_z}\gamma^\mathrm{X}_\mathrm{rec}$, and  $1/\gamma_\mathrm{exc}$  time between subsequent exciton capture events. $\tau_\mathrm{orient}$ depends thus linearly $\propto 1/\gamma_\mathrm{exc}$, but the large power (large $\gamma_\mathrm{exc}$) extrapolation is not towards zero, but the dark exciton lifetime multiplied by $1+1/p_f$.

These approximate analytical results agree with numerical simulations for $p_f=40$\%, see Fig.~\ref{fig6}(a). Note that the dependence of the orientation time on $1/\gamma_\mathrm{exc}$ disagrees with $\tau_\mathrm{orient}\propto1/\gamma_\mathrm{exc}$ measured in \cite{Goryca_PRL09}. This is because the orientation time is limited at high powers by the dark exciton lifetime. In Fig.~\ref{fig6}(b,c) we can see that only of  in a somewhat unrealistic case of $\epsilon=0.2$ and $A_h/A_e=2.5$ the dark X lifetime is short enough for $\tau_\mathrm{orient}$ to be proportional to  $1/\gamma_\mathrm{exc}$. In these Figures we can also see that the slope of $\tau_\mathrm{orient}(\tau_\mathrm{exc})$ dependence (in its linear regime) is independent of the QD parameters.

\begin{figure}
\centering
\includegraphics{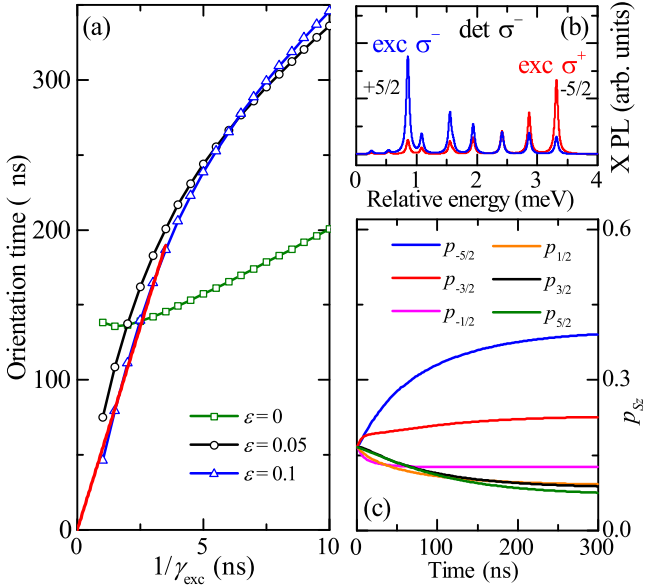}
\caption{(a) The orientation time of the Mn$^{2+}$ spin (determined by fitting the numerically-computed $p_{-5/2}(t)$ with an exponential curve) as a function of $1/\gamma_\mathrm{exc}$ for $\sigma^-$-polarized excitation and the following model parameters: $p_f=40$\%, $p_m=65$\%, $A_h=0.8$~meV, $A_e=0.16$~meV, $\delta_0=1.0$~meV, $\delta_1=0$, $1/\gamma^\mathrm{X}_{\mathrm{rec}}=400$~ps, $1/\gamma^\mathrm{2X}_{\mathrm{rec}}=250$~ps, and various strengths of the hh-lh mixing. The solid straight line represents the fit of a linear dependence $\tau_\mathrm{orient}\propto\tau_\mathrm{exc}$ to the curve corresponding to $\epsilon=0.1$ in the range of large excitation powers (i.e., $1/\gamma_\mathrm{exc}<3.5$~ns). (b),(c) Steady-state X PL spectra (b) and the time dependencies of $p_{S_z}$ (c) numerically computed for $\sigma^-$-polarized excitation, $p_m=65$~\%, $\epsilon=0.1$, and $1/\gamma_\mathrm{exc}=2$~ns.\label{fig7}}
\end{figure}

In principle, the exciton can also change its total spin due to the electron spin-flip taking place during the inter-dot transfer. While there is little experimental data on the efficiency of such process, we can theoretically consider its influence on the Mn$^{2+}$ spin orientation by introducing the $p_m \geq 50$\% parameter denoting the fraction of excitons that enter the Mn-containing dot with unflipped electron spin. 
Of course in the presence of both hole and electron spin relaxation the connection between the degree of circular polarization of absorbed light and the spin polarization of excitons coming into contact with the Mn$^{2+}$ spin is weakened, and the steady-state value of Mn$^{2+}$ spin polarization is expected to be lowered. It is less obvious that the characteristic timescale of orientation is also strongly affected --- in fact lengthened --- by presence of the additional spin relaxation channel. This feature arises due to presence of more than two Mn$^{2+}$ spin levels (i.e.~the fact that $S=5/2$), and it can be understood using a simple toy model, see Appendix \ref{Dodatek}. For the full model considered here we have to resort to numerical simulations, the results of which are shown in Fig.~\ref{fig7}(a) for $p_f=40$\% and $p_m=65$\% (typically observed \cite{Smolenski_PRB_2015_2X}). Comparison of the result for $\epsilon \! =\! 0.1$ with the one presented in Fig.~\ref{fig6}(a) shows how $p_m<100$\% increases the orientation time.

\begin{figure}
\centering
\includegraphics{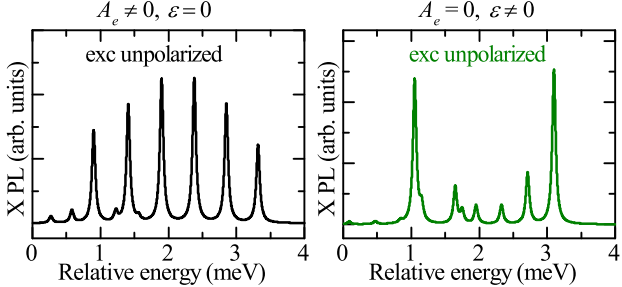}
\caption{Steady-state X PL spectra due to unpolarized excitation computed for $p_f=40$\%, $\delta_0=1.0$~meV, $\delta_1=0$, $1/\gamma^\mathrm{X}_{\mathrm{rec}}=400$~ps, $1/\gamma^\mathrm{2X}_{\mathrm{rec}}=250$~ps, and $1/\gamma_\mathrm{exc}=4$~ns. The PL spectrum from the panel (a) corresponds to the case of absence of hh-lh mixing (i.e., $A_h=0.8$~meV, $A_e=0.16$~meV, $\epsilon=0$), while the spectrum from the panel (b) corresponds to the absence of electron-Mn$^{2+}$ exchange interaction (i.e., $A_h=0.8$~meV, $A_e=0$, $\epsilon=0.1$). \label{fig8}}
\end{figure}

Finally let us revisit the question of PL spectrum shape for linearly polarized excitation. Two examples of the spectra for $p_f=40$\%  are shown in Fig.~\ref{fig8}: in Fig.~\ref{fig8}(a) only e-Mn$^{2+}$ interaction leads to brightening on the dark excitons (since $\epsilon \! =\! 0$), while in Fig.~\ref{fig8}(b) only the h-Mn$^{2+}$ interaction is relevant for brightening (since $A_e$ is put equal to zero). These cases correspond to the ones described in Sections \ref{subsec:onlyeMn} and \ref{subsec:onlyhMn}, respectively. Comparison with Figs.~\ref{fig1}(e) and \ref{fig3}(d) shows that in the currently discussed model the characters of the PL spectra for the two cases have traded places with respect to previously shown results. The primary reason for this effect is a difference in mixing of bright and dark excitonic states, which leads to the orientation of the Mn$^{2+}$ in both mechanisms. For instance, let us consider the case of $\epsilon \! =\! 0$. If the spin projection of the Mn$^{2+}$ ion is equal to $S_z$ and the bright $J_z=-1$ exciton is injected to the QD, the Mn$^{2+}$ spin is oriented owing to the mixing between the bright $|J_z=-1,S_z\rangle$ state and the dark $|J_z=-2,S_z+1\rangle$ state, the amplitude of which is equal to $F_{-2,-1,S_z}$. On the other hand, if the hole flips its spin during the inter-dot transfer, and the exciton is thus injected as a dark one with $J_z=+2$ (under $\sigma^-$-polarized excitation), the Mn$^{2+}$ spin orientation is then due to the mixing between $|J_z=+2,S_z\rangle$ and $|J_z=+1,S_z+1\rangle$ state, the amplitude of which is different and yields $F_{2,1,S_z+1}$. In fact, this amplitude is identical to the one appearing in the formula describing the rate of the Mn$^{2+}$ spin orientation from the $S_z+1$ state to $S_z$ under $\sigma^+$-polarized excitation (see Sec.~\ref{subsec:onlyeMn}). The above analysis clearly demonstrates that the spin orientation rate due to both considered mechanisms is an increasing function of \textit{the same} mixing amplitudes for \textit{opposite} circular polarizations of excitation. This finally leads to aforementioned, opposite characters of the X PL spectra under unpolarized excitation for both mechanisms.

\section{Comparison with the experiment and inclusion of Mn$^{2+}$ spin flipping during the energy relaxation of the exciton}
\label{seq:fullModel}

\begin{figure*}
\includegraphics{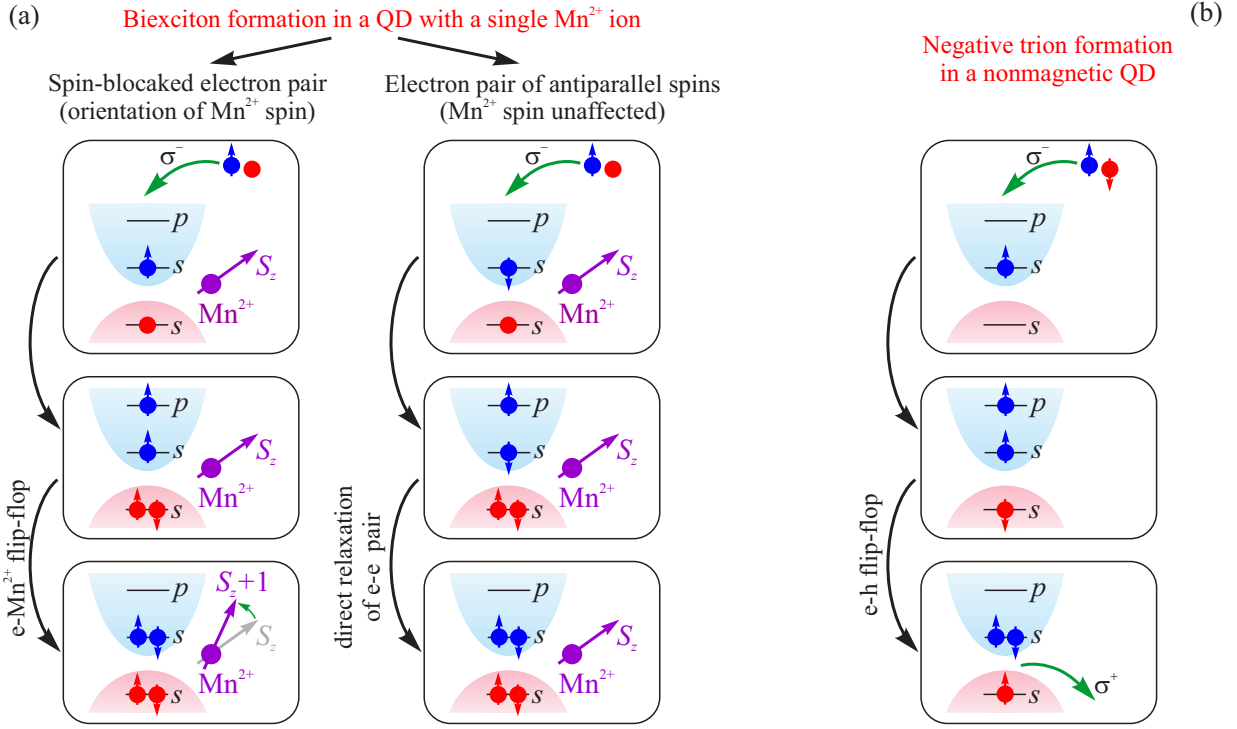}
\caption{(a) Scheme illustrating subsequent stages of the biexciton formation in a QD containing an individual Mn$^{2+}$ ion excited through an adjacent QD with $\sigma^-$-polarized light. Each of the two columns represents the situation, in which the spin of the electron injected to the QD is oriented either parallel or antiparallel to the spin of the second electron that is already present in the dot (and occupies its ground, $s$-shell state). (b) Analogous scheme presenting the formation process of a negatively charged trion in a nonmagnetic QD for the case, in which the relaxation of spin-blockaded electron pair is mediated by the flip-flop process with the spin of the hole residing on the $s$-shell.
\label{fig9}}
\end{figure*}

Let us assess now to what degree the previously discussed model of optical orientation can explain the results from Ref.~\onlinecite{Goryca_PRL09}, namely the direction of orientation (towards $S_z \! =\! \pm 5/2$ for $\sigma^{\mp}$ excitation), the typical orientation time of 20~ns --- 100~ns for $1/\gamma_\mathrm{exc}$ of 0.5 --- 3 ns, linear proportionality of both times (i.e.~$\tau_\mathrm{orient}\propto 1/\gamma_\mathrm{exc}$), and, most importantly, the fact that all these features appear to be generic for coupled double quantum dots, since they were observed for multiple QDs. 

For introduction of only bright excitons into the Mn-containing QD (Section \ref{sec:minimal_model}), the orientation direction agrees with experiment when  e-Mn$^{2+}$ flip-flop plays a dominant role, while it disagrees when hole-Mn$^{2+}$ flip-flop governs the dynamics. In presence of competing carrier-Mn$^{2+}$ exchange interactions, the net polarization sign depends on QD parameters. Finally, the typical orientation time is about an order of magnitude longer than the observed one.

The addition of processes of carrier spin relaxation during the exciton transfer (Section \ref{sec:dark_and_bright}) stabilizes the orientation direction (for $p_f \! \approx \! 40$\% this direction is in agreement with the observed one independently of QD parameters), and makes the orientation time shorter (i.e.~closer to the observed one). However, the power dependence of the orientation time is $\tau_\mathrm{orient}\propto  1/\gamma_\mathrm{exc} + \text{const.}$, with the offset controlled by the dark exciton lifetime. 

These results suggest that the explanation of experiments from \cite{Goryca_PRL09} requires an introduction of additional elements to the model. Since we have already considered essentially all the processes involving the Mn$^{2+}$ spin and the exciton in its orbital ground state, we conclude that the orientation observed in \cite{Goryca_PRL09} occurs (at least in part) due to Mn$^{2+}$ flipping processes that happen {\it during} the exciton transfer from one dot to the other. Full analysis of such processes is impossible due to lack of reliable information on the structure of excited states, and the nature and strength of Mn$^{2+}$ spin interaction with an exciton in one of such states. Only the interaction of Mn$^{2+}$ with an exciton in the first excited state  was characterized \cite{Smolenski_PRB_2015_x2m}. We will focus then on dynamics that occurs in the light-emitting dot just before the relaxation of the exciton into its orbital ground state. Interestingly, using the results obtained for nonmagnetic coupled QDs, we are able to propose a physical model that explains all the features of orientation observed in \cite{Goryca_PRL09}.

\begin{figure*}
\includegraphics{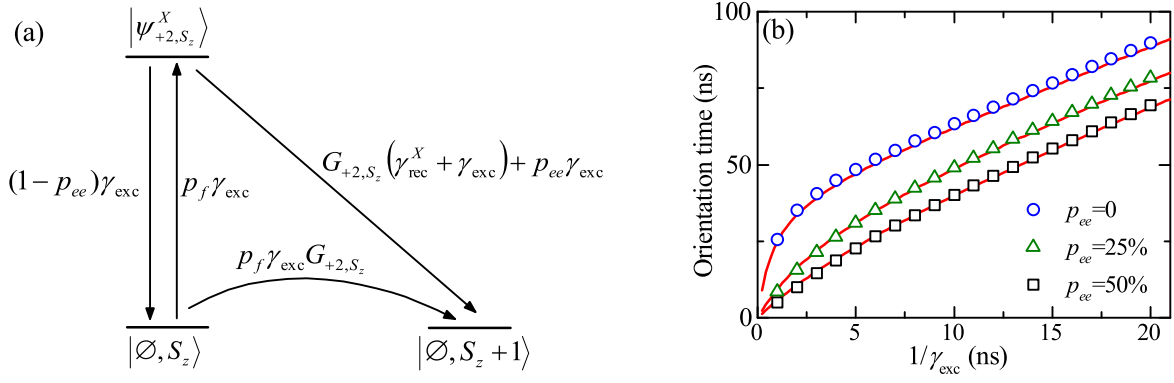}
\caption{(a) The simplified diagram of states (relevant for the case of $\gamma_\mathrm{exc}\ll\gamma^\mathrm{X}_{\mathrm{rec}}<\gamma^\mathrm{2X}_{\mathrm{rec}}$) illustrating both key mechanisms responsible for the Mn$^{2+}$ spin orientation: the one related to injection (and subsequent recombination) of the dark excitons, and the other one related to the electron-Mn$^{2+}$ flip-flop taking place during the formation of the biexciton. (b) The orientation time of the Mn$^{2+}$ spin (determined by fitting a numerically-computed time dependence of $p_{-5/2}$ with an exponential curve) as a function of the inverse (relative) excitation power $1/\gamma_\mathrm{exc}$. The calculations were performed for $\sigma^-$ polarization of excitation, $p_f=40$\% and the QD parameters provided in the caption of Fig.~\ref{fig5}. Each out of three presented series of points corresponds to a different electron-Mn$^{2+} $flip-flop probability $p_{ee}$. The solid lines represent the results of simplified analytical formula given by Eq. (\ref{Eq:allMnrate}).
\label{fig10}}
\end{figure*}

In the new scenario of dynamics we focus on the process of creation of biexciton in the orbital ground state of Mn-containing QD. We assume that one exciton is already present in this state, and the second one is relaxing towards it. Time-resolved measurements on nonmagnetic pairs of coupled QDs \cite{Smolenski_PRB_2015_2X} show, that the holes from the two excitons relax in $<\! 15$ ps (independently of their spins) to the lowest energy state, in which they form a spin singlet decoupled from the electrons (and the Mn$^{2+}$ spin in the case of magnetic QD). The relaxation of the two electrons is, on the other hand, much slower: it is preceded by creation of an excited state in which one electron occupies the $s$ shell, while the other resides in the $p$ shell. The rate of subsequent energy relaxation depends strongly on the relative spin orientation of the electrons. In nonmagnetic QDs, a pair with antiparallel spins relaxes in $\approx \! 30$ ps, while a pair with parallel spins relaxes in at least twice as long a time $\approx \! 80$ ps. The presence of Mn$^{2+}$ ion can significantly increase the latter process by lifting the spin blockade: e-Mn$^{2+}$ exchange interaction (dominated by exchange with the electron in the $s$ shell) allows for simultaneous flipping of one of the electron spins and the Mn$^{2+}$ spin. Such a flip-flop is shown schematically in Fig.~\ref{fig9}(a). In this case the creation of biexcitons out of two excitons having parallel spin electrons is accompanied by Mn$^{2+}$ spin change. Note that since the electron spin is almost perfectly conserved during the exciton transfer, the probability of creation of spin-blocked electron pair is high under circularly polarized excitation. The spin of the electrons is then $\pm1$ for $\sigma^\mp$ excitation, so that the electron-Mn$^{2+}$ flip-flop results in change of spin from $S_z$ to $S_z\pm1$, in agreement with observations. 

A completely analogous process takes place during creation of negative trion (X$^{-}$) in a nonmagnetic dot \cite{Cortez_PRL02,Kazimierczuk_PRB09}. There, the spin blockaded electrons relax via flip-flop of one of them with the hole from the $s$ shall (see Fig.~\ref{fig9}(b)), which is enabled by an anisotropic electron-hole exchange. This process leads to optical orientation of X$^{-}$ being opposite to the ``natural'' one under circularly polarized excitation, a feature observed in all self-assembled QDs (and which persists up to $\sim\! 20$ T magnetic fields in CdTe QDs). The understanding of the apparent effectiveness of this process has been achieved only recently~\cite{Benny_PRB_2014}, by noticing the crucial role of electron-LO phonon Fr\"ohlih interaction that leads to renormalization of singlet-triplet splitting, and consequently to enhancement of their mixing by exchange interaction. The key observation is that the relevant physics is independent of the nature of the electron flip --- if the flip-flop with the hole for X$^{-}$ is efficient, then the flip-flop with Mn$^{2+}$ during biexciton creation will be efficient as well.

\begin{figure*}
\includegraphics{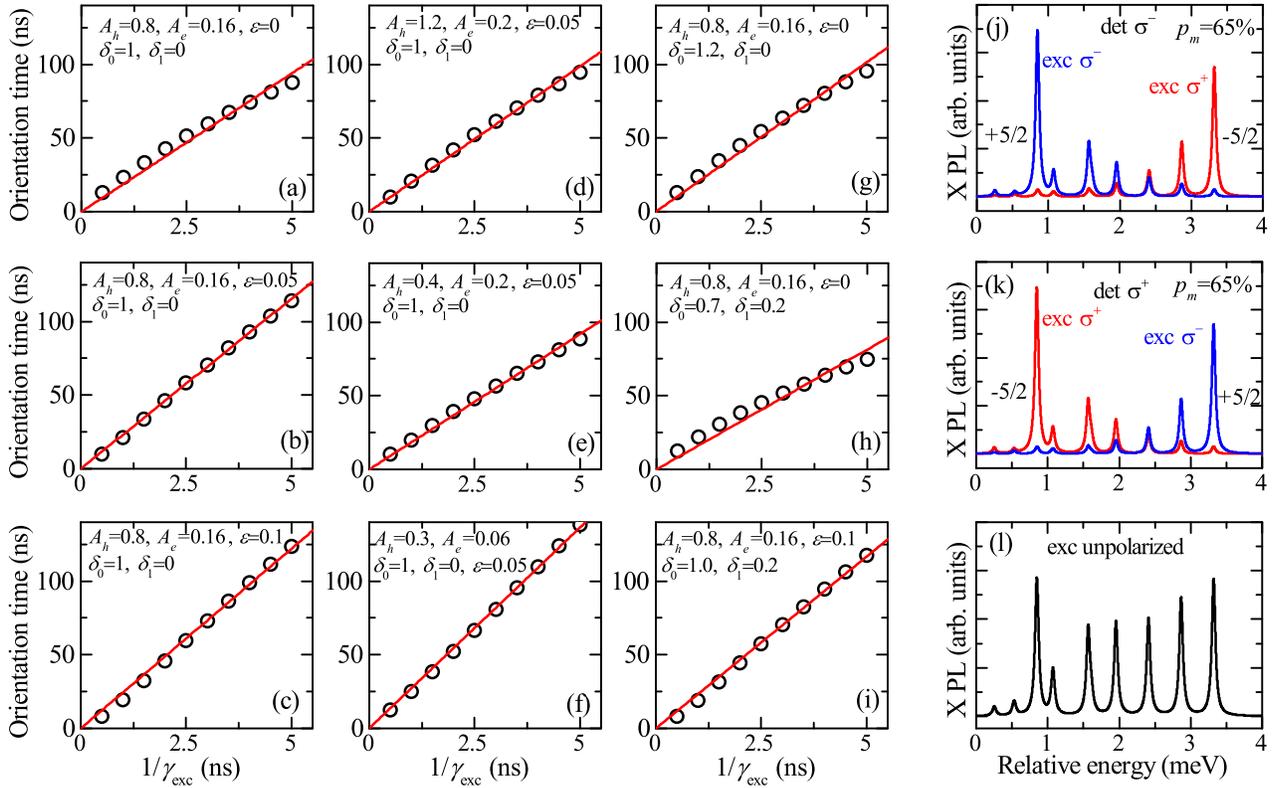}
\caption{(a-i) Dependencies of the Mn$^{2+}$ spin orientation time on $1/\gamma_\mathrm{exc}$ determined based on an exponential fit of numerically-computed $p_{-5/2}(t)$. The calculations were carried out for $\sigma^-$-polarized excitation (with the efficiency of the polarization transfer $p_m=65$\%), and for $p_f=40$\%, $p_{ee}=50$\%, $1/\gamma^\mathrm{X}_{\mathrm{rec}}=400$~ps, $1/\gamma^\mathrm{2X}_{\mathrm{rec}}=250$~ps. All the other model parameters are given in the figure (the exchange energies are expressed in meV). The solid lines represent the fitted linear dependencies of the form $\tau_\mathrm{orient}\propto1/\gamma_\mathrm{exc}$. (j-l) Steady-state X PL spectra computed for various polarizations of excitation and detection for $A_h=0.8$~meV, $A_e=0.16$~meV, $\epsilon=0.1$, $\delta_0=1.0$~meV, $\delta_1=0.2$~meV, and $1/\gamma_\mathrm{exc}=2$~ns.
\label{fig11}}
\end{figure*}

We include the above mechanism by modifying the biexciton creation rate: we define the probability $p_{ee}$ of biexciton creation via lifting of electron spin blockade by e-Mn$^{2+}$ flip-flop. For simplicity we assume that $p_{ee}$ is independent of the Mn$^{2+}$ spin, as long as the flip-flop is actually possible. The rate of creation of biexciton in 
 $|2\mathrm{X}, S_z\rangle$  state under $\sigma^-$ excitation is given by
\begin{widetext}
\begin{multline}
\gamma(|\psi^X_i\rangle\rightarrow|2\mathrm{X}, S_z\rangle)=\gamma_{\mathrm{exc}}\sum_{J_z=+1,-2}|\langle J_z,S_z|\psi^X_i\rangle|^2+\\+\gamma_{\mathrm{exc}}\sum_{J_z=-1,+2}\left[p_{ee}|\langle J_z,S_z-1|\psi^X_i\rangle|^2+(1-p_{ee})|\langle J_z,S_z|\psi^X_i\rangle|^2\right] \,\, ,
\end{multline}
\end{widetext}
for $S_z>-5/2$, and by $\gamma(|\psi^X_i\rangle\rightarrow|2\mathrm{X}, S_z\rangle)=\gamma_{\mathrm{exc}}\sum_{J_z}|\langle J_z,S_z|\psi^X_i\rangle|^2$ for $S_z=-5/2$. This mechanism is effective only for large enough excitation power, for which the time between appearance of one exciton and the second one is shorter than the exciton radiative lifetime. Consequently, high  Mn$^{2+}$ spin orientation rate due to the flip-flop process of spin blockade lifting can be achieved when many dark excitons enter the light-emitting QD (i.e.~when $p_f$ is large). This means that the currently discussed mechanism is working hand in hand with the mechanism described in Sec.~\ref{sec:dark_and_bright}. We can understand it by modifying the transitions scheme from Fig.~\ref{fig5}(e). We need to modify the transitions corresponding to creation of biexciton out of a mostly dark $|\psi^X_{+2,S_z}\rangle$ state: the probability of biexciton creation in  $|2\mathrm{X},S_z\rangle$ state is lowered from  $\gamma_\mathrm{exc}$ to $(1-p_{ee})\gamma_\mathrm{exc}$, and there appears a possibility of creation of biexciton in  $|2\mathrm{X},S_z+1\rangle$ state with $p_{ee}\gamma_\mathrm{exc}$ rate. Under the approximation of $\gamma_\mathrm{exc}\ll\gamma^\mathrm{X}_{\mathrm{rec}}<\gamma^\mathrm{2X}_{\mathrm{rec}}$ we arrive at a simplified transition scheme shown in Fig.~\ref{fig10}(a), and at an approximate formula for the Mn$^{2+}$ orientation rate:
\begin{eqnarray}
\gamma_\mathrm{orient}&&(S_z\xrightarrow{\sigma^{-}} S_z+1)\approx \nonumber \\ &&p_{ee}p_f\gamma_\mathrm{exc}\frac{\gamma_\mathrm{exc}}{(1+p_f)\gamma_\mathrm{exc}+G_{2,S_z}\gamma^\mathrm{X}_\mathrm{rec}}+\nonumber \\
&&+G_{2,S_z}\gamma_\mathrm{exc}\frac{p_f\left(\gamma^\mathrm{X}_\mathrm{rec}+2\gamma_\mathrm{exc}\right)}{(1+p_f)\gamma_\mathrm{exc}+G_{2,S_z}\gamma^\mathrm{X}_\mathrm{rec}} \,\, .
\label{Eq:allMnrate}
\end{eqnarray}
This formula is a sum of two contributions, the first of which comes from spin blockade lifting by electron-Mn$^{2+}$ flip-flop, while the second one is the orientation rate $\gamma_d(S_z\xrightarrow{\sigma^{-}} S_z+1)$ caused by appearance and recombination of dark excitons, cf.~Eq.~(\ref{Eq:dMnrate}). The ratio of the two contributions is $\approx \! p_{ee}\gamma_\mathrm{exc}/G_{2,S_z}\gamma^\mathrm{X}_\mathrm{rec}$, so that for large $p_{ee}$ and $\tau_\mathrm{exc} \! = \!1/\gamma_\mathrm{exc}$ time shorter than the dark exciton lifetime $G_{2,S_z}\gamma^\mathrm{X}_\mathrm{rec}$, the flip-flop mechanism of Mn$^{2+}$ spin orientation that is the dominant one. Consequently, for large excitation powers the orientation rate is proportional to $\gamma_\mathrm{exc}$, in agreement with experiments \cite{Goryca_PRL09}.

Numerical simulations for $p_f=40$\% and various $p_{ee}$ confirm the above approximate result, see Fig.~\ref{fig10}(b). With increasing $p_{ee}$, the offset of the linear dependence of  $\tau_\mathrm{orient}$ on $\tau_\mathrm{exc}$ decreases, and for $p_{ee}=50$\% the orientation time becomes practically proportional to $\tau_\mathrm{exc}$. Furthermore, as illustrated in Fig.~\ref{fig11}(a)-(i), this result does not require any fine-tuning of the QD parameters. We have used $p_m=65$\% and $1/\gamma_\mathrm{exc}<5$~ns in these simulations, in agreement with the experiment \cite{Goryca_PRL09}. The proportionality factor between $\tau_\mathrm{orient}$ and $\tau_\mathrm{exc}$ is between $15$ and $30$ for a wide range of used parameters, showing that it is weakly dependent on strenght and character (e.g.~the presence of hh-lh mixing) of carrier-Mn$^{2+}$ exchange interactions. This is a simple consequence of Eq.~(\ref{Eq:allMnrate}), which for large excitation power  gives  
\begin{equation} 
\gamma_\mathrm{orient}(S_z\xrightarrow{\sigma^{-}} S_z+1)\approx[p_{ee}p_f/(1+p_f)]\gamma_\mathrm{exc} \,\, . 
\end{equation}
From this we get that the probability of Mn$^{2+}$ spin flip from $S_z$ to $S_z\pm1$ per one polarized excitons injected into the QD depends only on $p_{ee}$ and $p_{f}$, and for $p_{ee}=50$\% and $p_f=40$\% we have $p_{0} \! \approx \! p_{ee}p_f/(1+p_f)\approx0.14$, which is close to the  $p_0\! =\! 0.1$ estimated in \cite{Goryca_PRL09} (note that proportionality factors from Figs.~\ref{fig11}(a-i) are between $2$ and $4$ times larger than $p_{0}$ due to the assumption of imperfect polarization transfer, see Appendix). Finally, let us not that the calculated steady state  PL spectrum of the neutral exciton, shown in Fig.~\ref{fig11}(j,k,l), bears close resemblance to the observed spectra \cite{Goryca_PRL09,Goryca_PRB_2015}.

\section{Summary}
We have discussed a broad range of models of Mn$^{2+}$ spin orientation due to quasi-resonant injection of spin-polarized excitons into a quantum dot containing a single Mn$^{2+}$ spin. All of these models were based on a general assumption that the quantum dot remains in a neutral charge state, and that it may be occupied by an exciton or a biexciton. The spin orientation process was considered within a frame of a simple rate-equation model describing the changes of the Mn$^{2+}$ spin state induced by the $s,p$-$d$ exchange interaction with optically created excitons. Our analysis revealed that, depending on the specific assumptions, such a model predicts vastly different orientation rates or even different directions of the orientation process. The various discussed mechanisms could be used to understand future experiments on analogous systems involving QDs containing single magnetic ions other than Mn$^{2+}$. The sensitivity of the optical orientation process to various parameters characterizing the excitonic states, their coupling with the localized spin, and the relaxation processes of all the involved spins, could be used to indirectly access these parameters that are most often inaccessible to direct measurement (e.g.~carrier spin relaxation times, that are typically much longer than the exciton recombination rate, but the magnitude of which influences the Mn$^{2+}$ photo-orientation time, that is a much more easily measurable quantity).

Building upon these findings we were able to quantitatively reproduce the experimental results from Ref.~\onlinecite{Goryca_PRL09}. The two key ingredients were: (1) sufficiently high influx of the dark excitons (with spin-flipped hole), and (2) a possibility of an efficient flip-flop process between the spins of the electron and the Mn$^{2+}$ ion, assisting the biexciton formation.

\section*{Acknowledgements}
We thank T. Kazimierczuk and M. Goryca for fruitful discussions. This work was supported by the Polish National Science Center (NCN) Project No. DEC-2011/02/A/ST3/00131 and by the Polish Ministry of Science and Higher Education in years 2012-2016 as a research grant ''Diamentowy Grant''. T.S. was supported by the Polish National Science Centre through PhD scholarship Grant No. DEC-2016/20/T/ST3/00028. L.C. acknowledges the support from Polish National Science Centre through Grant No. DEC-2014/13/B/ST3/04603.

\appendix
\section{Dependence of the Mn$^{2+}$ spin orientation rate on the efficiency of the exciton polarization transfer in a simplified model}
\label{Dodatek}

In order to elucidate the reason for the increase of the Mn$^{2+}$ orientation time with decreasing efficiency of the exciton polarization transfer between the coupled QDs, it is instructive to employ the simplest \textit{rate-equation} model, which was originally proposed in Ref.~\onlinecite{Goryca_PRL09}. In the generalized version of this model we consider $n>1$ different spin states ($6$ in the case of the Mn$^{2+}$ ion), and label their occupation probabilities by $p_i$. We assume that initially, i.e., at $t=0$, each of these probabilities is the same (i.e., $p_i(t=0)=1/n$). The optical orientation process is then described by introducing transitions between adjacent states. More specifically, we assume that during the presence of $J_z=\pm1$ exciton the $i$ state can be transferred to $i\pm1$ state with the rate of $\gamma_\mathrm{orient}$ (being dependent on the excitation power). The limited efficiency of the polarization transfer is accounted for by introducing the parameter $\alpha>0.5$, which corresponds to a probability that the exciton $J_z=\pm1$ is injected to the dot under $\sigma^\pm$-polarized excitation. Within these assumptions we obtain the following set of rate-equations relevant for the excitation with $\sigma^-$ polarization (which are schematically illustrated in Fig.~\ref{fig_appendix}(a)):
\begin{widetext}
\begin{equation}
\frac{dp_i}{dt} = \begin{cases} -\alpha\gamma_\mathrm{orient}p_1+(1-\alpha)\gamma_\mathrm{orient}p_2, & \mbox{if } i=1, \\ -\gamma_\mathrm{orient}p_i+\alpha\gamma_\mathrm{orient}p_{i-1}+(1-\alpha)\gamma_\mathrm{orient}p_{i+1}, & \mbox{if } 1<i<n, \\
-(1-\alpha)\gamma_\mathrm{orient}p_n+\alpha\gamma_\mathrm{orient}p_{n-1}, & \mbox{if } i=n. \end{cases}
\label{Eq:toyratemodel}
\end{equation}

Let us now consider how fast the leftmost $i=1$ state is being depleted. In an idealized case of $\alpha=1$ (i.e., 100\% efficiency of the polarization transfer), this process is simply described by $\frac{dp_1}{dt}=-\gamma_\mathrm{orient}p_1$, which shows the depletion (i.e., orientation) rate to be equal to $\gamma_\mathrm{orient}$, independently of the number $n$ of considered states. Analytical solution of the aforementioned set of rate-equations shows this result to be unaltered in the general case of $\alpha<1$ as long as there are only $n=2$ states in the model. However, for larger number $n\ge 3$ of states the situation turns out to be completely different. In order to exemplify this, let us consider the particular case of $n=3$, for which the occupation probability of the leftmost state is given by
\begin{equation}
p_1(t)=\frac{(1-\alpha)^2}{1-\beta}+\frac{(2\alpha-1)(2-\alpha)}{3(1-\beta)}e^{-\gamma_\mathrm{orient}t}\left[\cosh\left(\sqrt{\beta}\gamma_\mathrm{orient}t\right)+\frac{1+\alpha}{2-\alpha}\sqrt{\frac{1-\alpha}{\alpha}}   \sinh\left(\sqrt{\beta}\gamma_\mathrm{orient}t\right)\right],
\label{Eq:p1_3states_toy}
\end{equation}
where $\beta=\alpha(1-\alpha)$. This time dependence exhibits multi-exponential character, but, analogously as in Sec.~\ref{sec:dark_and_bright}, it can be approximated with an exponential dependence of the form $p_1(t)\approx p_1(\infty)+[p_1(0)-p_1(\infty)]\exp(-\gamma_{n=3}(\alpha)t)$, where $\gamma_{n=3}(\alpha)$ represents the effective depletion rate of the considered state. The value of $\gamma_{n=3}(\alpha)$ may be thus calculated based on the time integral of $p_1(t)-p_1(\infty)$ as
\begin{figure*}
\includegraphics{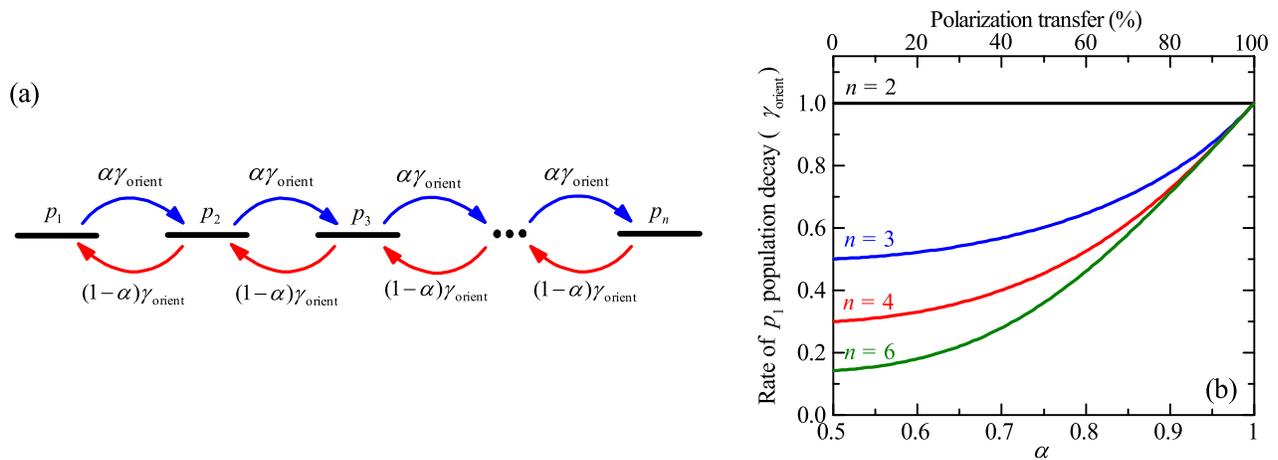}
\caption{(a) Scheme of states and transitions included in the simplified model of the optical orientation process. (b) Dependencies of the depletion time of the leftmost state ($i=1$) on the polarization transfer efficiency $\alpha$ calculated for various number $n$ of states considered in the model.
\label{fig_appendix}}
\end{figure*}
\begin{equation}
\gamma_{n=3}(\alpha)=\left(\frac{1}{p_1(0)-p_1(\infty)}\int_0^\infty \left[p_1(t)-p_1(\infty)\right]dt\right)^{-1}=\gamma_\mathrm{orient}\frac{1+(1-\alpha)^3}{1+3(1-\alpha)-(1-\alpha)^2}.
\end{equation}
\end{widetext}
Fig.~\ref{fig_appendix}(b) presents the dependence of the above calculated orientation rate $\gamma_{n=3}$ on $\alpha$. As seen, the considered rate clearly decreases for lower values of $\alpha$, and turns out to be almost two times smaller than $\gamma_\mathrm{orient}$ for $\alpha\approx0.7$ known from the experiment~\cite{Goryca_PRL09}. Even more pronounced difference between the these two rates is obtained for larger number of states, which is revealed by $\gamma_{n=4}(\alpha)$ and $\gamma_{n=6}(\alpha)$ dependencies shown in Fig.~\ref{fig_appendix}(b). In particular, the value of $\gamma_{n=6}(\alpha\approx0.7)$ for $n=6$ (i.e., the case of the Mn$^{2+}$ ion) is approximately four times smaller than $\gamma_\mathrm{orient}$. This result unequivocally confirms that the Mn$^{2+}$ spin orientation dynamics (measured experimentally as the depletion rate of state with an extreme spin projection) is substantially influenced by the efficiency of the excitonic polarization transfer.

\end{document}